\documentclass[11pt]{article}
\usepackage{mathptmx}
\usepackage{helvet}
\usepackage{courier}
\usepackage{type1cm}
\usepackage{makeidx}
\usepackage{graphicx}
\usepackage{multicol}
\usepackage[bottom]{footmisc}

\begin{document}

%\title{A Gravitational Aharonov-Bohm Effect, and its Connection to Parametric Oscillators and
%Gravitational Radiation}
% Use 
%\titlerunning{Gravitational AB Effect, and its Connection to Parametric Oscillators and GR radiation}
% for an abbreviated version of

\begin{center}\huge{A Gravitational Aharonov-Bohm Effect,\\and its Connection to Parametric Oscillators and Gravitational Radiation} \\
\vspace{0.1in}

\large{Raymond Y. Chiao$^1$, Robert Haun$^2$, Nader Inan$^2$, Bong-Soo Kang$^2$, \\ Luis A. Martinez$^2$, Stephen J.
Minter$^3$, Gerardo Mu\~noz$^4$, and Douglas Singleton$^{4,5}$} \\ \vspace*{0.2in} \scriptsize{$^1$University of California, Merced,
Schools of Natural Sciences and Engineering,
P.O. Box 2039, Merced, CA
95344, USA \\
Corresponding author, email: rchiao@ucmerced.edu  \\
\vspace{0.05in} $^2$University of California, Merced, School of Natural
Sciences, P.O. Box 2039, Merced, CA 95344, USA \\
\vspace{0.05in} $^3$Vienna Center for Quantum Science and Technology,
Faculty of Physics, 
 University of Vienna, Boltzmanngasse 5, A-1090 Vienna, Austria \\ 
\vspace{0.05in} $^4$California State University, Fresno, CA 93740, USA
\vspace{0.05in} $^5$ Department of Physics, Institut Teknologi Bandung, Indonesia}
% Use
% \authorrunning{Chiao, Haun, Inan, Kang, Martinez, Minter, Mu\~noz, and Singleton}
%  for an abbreviated version of
%\institute{}
\end{center}


\abstract{A thought experiment is proposed to demonstrate the
existence of a gravitational, vector Aharonov-Bohm effect. We begin the
analysis starting from four Maxwell-like equations for weak gravitational
fields interacting with slowly moving matter. A connection is made between
the gravitational, vector Aharonov-Bohm effect and the principle of local
gauge invariance for nonrelativistic quantum matter interacting with weak
gravitational fields. The compensating vector fields that are necessitated
by this local gauge principle are shown to be incorporated by the DeWitt
minimal coupling rule. The\ nonrelativistic Hamiltonian for weak,
time-independent fields interacting with quantum matter is then extended to
time-dependent fields, and applied to problem of the interaction of
radiation with macroscopically coherent quantum systems, including the
problem of gravitational radiation interacting with superconductors. But
first we examine the interaction of EM radiation with superconductors in a
parametric oscillator consisting of a superconducting wire placed at the
center of a high $Q$ superconducting cavity driven by pump microwaves. Some
room-temperature data will be presented demonstrating the splitting of a
single microwave cavity resonance into a spectral doublet due to the
insertion of a central wire. This would represent an \emph{unseparated} kind
of parametric oscillator, in which the signal and idler waves would occupy
the same volume of space. We then propose a \emph{separated} parametric
oscillator experiment, in which the signal and idler waves are generated in
two disjoint regions of space, which are separated from each other by means
of an impermeable superconducting membrane. We find that the threshold for
parametric oscillation for EM microwave generation is much lower for the
separated configuration than the unseparated one, which then leads to an
observable dynamical Casimir effect. We speculate that a separated
parametric oscillator for generating coherent GR microwaves could also be built.
}

%\abstract{Each chapter should be preceded by an abstract (10--15 lines long) that summarizes the content. The abstract will appear \textit{online} at \url{www.SpringerLink.com} and be available with unrestricted access. This allows unregistered users to read the abstract as a teaser for the complete chapter. As a general rule the abstracts will not appear in the printed version of your book unless it is the style of your particular book or that of the series to which your book belongs.\newline\indent
%Please use the 'starred' version of the new Springer \texttt{abstract} command for typesetting the text of the online abstracts (cf. source file of this chapter template \texttt{abstract}) and include them with the source files of your manuscript. Use the plain \texttt{abstract} command if the abstract is also to appear in the printed version of the book.}

\begin{figure}[h]
%\sidecaption[t]
\centering
\includegraphics[angle=0,width=.5\textwidth]{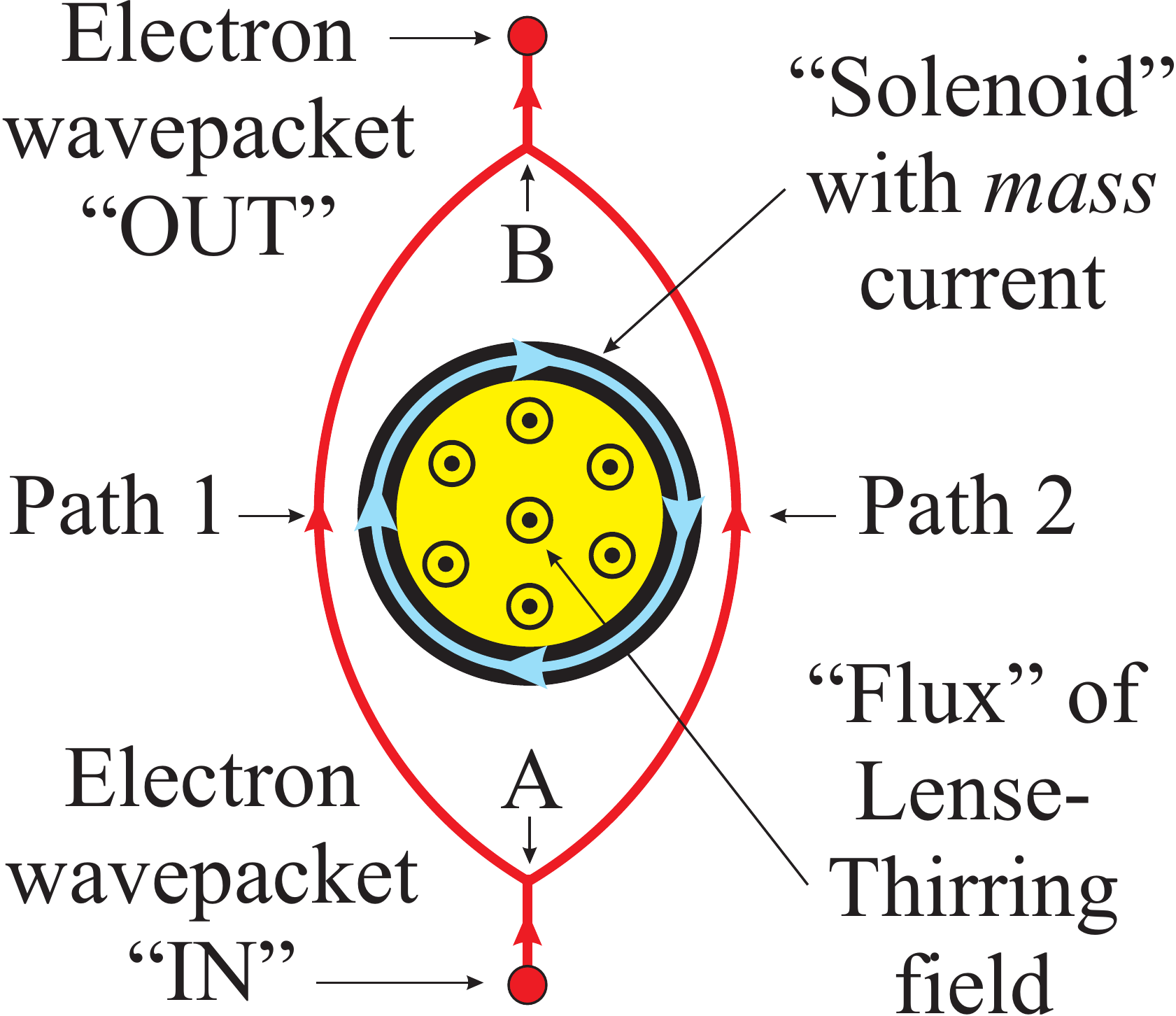} 
%\label{fig:REFERENCE}
\caption{(Color online) Sketch of a gravitational Aharonov-Bohm effect. A
\textquotedblleft solenoid\textquotedblright\ with circulating mass currents
(in blue), produces \textquotedblleft flux\textquotedblright\ (black dots)
of a certain \textquotedblleft gravito-magnetic\textquotedblright\ field
(the Lense-Thirring field) in its interior (in yellow). In its exterior (in
white),\ this field is zero. Nevertheless, an electron wave packet (in red),
which is split at point A to go around the \textquotedblleft
solenoid\textquotedblright\ via paths 1 and 2, and then recombined at point
B, will exhibit an Aharonov-Bohm fringe shift.}
\end{figure}

\section{Introduction}

In this paper we discuss three apparently distinct phenomena: The
gravitational Aharonov-Bohm effect, the dynamical Casimir effect arising
from parametric oscillations, and gravitational waves. The first of these
phenomena is simply the gravitational version of the electromagnetic
Aharonov-Bohm effect. There has been recent interest in the gravitational
version of the \emph{scalar} Aharonov-Bohm effect \cite{zeilinger}. Here we
will discuss the gravitational version of the \emph{vector} Aharonov-Bohm
effect. In the second phenomenon, i.e., the dynamical Casimir effect, we
propose a possible experiment in which\ photons could be \textquotedblleft
pumped out of the vacuum\textquotedblright\ via a vibrating superconducting
(SC) \textquotedblleft membrane\textquotedblright\ considered as a
parametric oscillator. Finally, we discuss gravitational waves. We speculate
that a gravitational version of \textquotedblleft pumping gravitons out of
the vacuum\textquotedblright\ via parametric amplification, and above
threshold, parametric oscillation, might be possible. The analog of a laser
for gravitational waves could thus be constructed. The concept that links
these three phenomena together is the use of the DeWitt minimal coupling
rule, whereby particles are coupled to both the electromagnetic and the
gravitational vector potential.

\section{Gravitational Aharonov-Bohm effect}

A gravitational analog of the vector Aharonov-Bohm effect is depicted in
Figure 1 \cite{zeilinger}\cite{A&C}\cite{harris}. Aharonov-Bohm (AB) interference can
occur when an incoming single-electron wavepacket is split at point A by
means of a beam splitter into two partial waves traveling along paths 1 and
2, respectively, that go around the outside of a \textquotedblleft
solenoid\textquotedblright\ which contains circulating \emph{mass} currents
(indicated by the blue arrows).

For instance, such mass currents in a cylindrical superconducting (SC) mass
shell (indicated in black in Figure 1), could be produced by rotating the
shell at a constant angular frequency around its cylindrical axis. The two
partial waves could then be recombined into an outgoing single-electron
wavepacket at point B by means of another beam splitter. Just like in the
purely electromagnetic AB effect, the \textquotedblleft
flux\textquotedblright\ of a certain gravito-magnetic field (i.e., the
Lense-Thirring field) would be confined entirely to the interior region
(indicated in yellow in Figure 1) of the \textquotedblleft
solenoid,\textquotedblright\ and would vanish at all points in its exterior
(indicated in white). There results a quantum mechanical AB fringe shift due
to the \textquotedblleft flux\textquotedblright\ that is observable at point
B, which cannot be explained classically.

To understand the thought experiment pictured in Figure 1, we begin from
Einstein's field equations, which, in the limit of weak gravitational fields near slowly moving matter (i.e., in the vicinity of nonrelativistic masses), become the
following set of four Maxwell-like equations \cite{BCT}: 
\begin{equation}
\mathbf{\nabla \cdot E}_{g}=-\frac{\rho _{g}}{\varepsilon _{g}}
\label{Maxwell-like-eq-1}
\end{equation}%
\begin{equation}
\mathbf{\nabla \times E}_{g}=-\frac{\partial \mathbf{B}_{g}}{%
\partial t}  \label{Maxwell-like-eq-2}
\end{equation}%
\begin{equation}
\mathbf{\nabla \cdot B}_{g}=0  \label{Maxwell-like-eq-3}
\end{equation}%
\begin{equation}
\mathbf{\nabla \times B}_{g}=4\mu _{g}\left( -\mathbf{j}_{g}+\varepsilon _{g}%
\frac{\partial \mathbf{E}_{g}}{\partial t}\right)  \label{Maxwell-like-eq-4}
\end{equation}%
where the gravitational analog $\varepsilon _{g}$ of the electric
permittivity $\varepsilon _{0}$ of free space is \cite{newtonian limit}%
\begin{equation}
\varepsilon _{g}=\frac{1}{4\pi G}=1.19 \times 10^{9} \textrm{ SI units} 
\label{epsilon_G}
\end{equation}%
and where the gravitational analog $\mu _{g}$ of the magnetic permeability $%
\mu _{0}$ of free space is \cite{mu_G}%
\begin{equation}
\mu _{g}=\frac{4\pi G}{c^{2}}=9.31\times 10^{-27} \textrm{ SI units}
\label{mu_G}
\end{equation}%
Here $G=6.67\times 10^{-11}$ SI units is Newton's constant, and $%
c=3.00\times 10^{8}~$m s$^{-1}$ in SI units is the vacuum speed of light.

In the four Maxwell-like equations, (\ref{Maxwell-like-eq-1}) $-$ (\ref%
{Maxwell-like-eq-4}), the electric-like field $\mathbf{E}_{g}$ is the \emph{%
gravito-electric} field, (i.e., the local acceleration $\mathbf{g}$ of a
freely falling test particle), which could be produced by the mass density $%
\rho _{g}$ of nearby matter, via (\ref{Maxwell-like-eq-1}). Likewise, the
magnetic-like field $\mathbf{B}_{g}$ is the \emph{gravito-magnetic} field,
which could be produced by the mass current density $\mathbf{j}_{g}$ of
nearby nonrelativistically moving matter, and also by the gravitational
analog of the Maxwell displacement current density $\varepsilon _{g}\partial 
\mathbf{E}_{g}/\partial t$, via (\ref{Maxwell-like-eq-4}). In the case of
nearby \emph{stationary} nonrelativistic mass currents, $\mathbf{B}_{g}$ can
be identified with the Lense-Thirring field \cite{Wald} that is generated by
these currents.

A nonrelativistic test particle with a mass $m$ moves in the presence of the
weak fields $\mathbf{E}_{g}$ and $\mathbf{B}_{g}$\ in accordance with the
Lorentz-like force law \cite{BCT}%
\begin{equation}
\mathbf{F}=m\frac{d\mathbf{v}}{dt}=m(\mathbf{E}_{g}+\mathbf{v\times B}_{g})
\label{Lorentz-like force law}
\end{equation}%
where $m$ is the mass of the test particle and $\mathbf{v}$ is its velocity (with $v<<c$).

To understand the experiment pictured in Figure 1, we only need the \emph{%
stationary} version of (\ref{Maxwell-like-eq-4}), i.e., the gravitational
analog of Ampere's law%
\begin{equation}
\mathbf{\nabla \times B}_{g}=-4\mu _{g}\mathbf{j}_{g}
\end{equation}%
where $\mathbf{j}_{g}$ and $\mathbf{B}_{g}$ do not depend on time. This
implies the following gravitational analog of Ampere's circuital law:%
\begin{equation}
\oint\limits_{C}\mathbf{B}_{g}\cdot d\mathbf{l}=-4\mu _{g}\left(
I_{g}\right) _{\textrm{enc}}
\end{equation}%
where $d\mathbf{l}$ is a line element of an arbitrary closed curve $C$, and $%
\left( I_{g}\right) _{\textrm{enc}}$ is the mass current which is enclosed by $C$.
Applying this Ampere's circuital law to the \textquotedblleft
solenoid\textquotedblright\ of Figure 1, which could be a uniformly rotating
SC cylindrical mass shell, and using an appropriately chosen closed curve $C$%
, one concludes that the $\mathbf{B}_{g}$ field in the interior to the
\textquotedblleft solenoid\textquotedblright\ is a uniform field pointing
along the cylindrical axis, and that it has a constant magnitude%
\begin{equation}
B_{g}=4\mu _{g}I_{g}^{\prime }
\end{equation}%
where $I_{g}^{\prime }$ is the mass current per unit length of the
\textquotedblleft solenoid\textquotedblright\ flowing around the
circumference of the rotating cylindrical mass shell. Furthermore, by
another appropriate choice of the closed curve $C$, one concludes that
everywhere outside the \textquotedblleft solenoid,\textquotedblright\ it is
the case that%
\begin{equation}
B_{g}=0
\end{equation}%
i.e., that the Lense-Thirring field vanishes everywhere exterior to the
\textquotedblleft solenoid.\textquotedblright\ This is analogous to the fact
that the magnetic field vanishes at all points outside of an electromagnetic
solenoid. Hence it follows from the Lorentz-like force law (\ref%
{Lorentz-like force law}) that although the electron in Figure 1 experiences
a \emph{radial} classical gravitational force due to the mass of the
\textquotedblleft solenoid,\textquotedblright\ it could never have
experienced any \emph{azimuthal}\ classical gravitational force on its way
from point A to point B via either path 1 or path 2 that could have caused
the AB phase shift.

Put differently, if one thinks of the \textquotedblleft
solenoid\textquotedblright\ as a rotating SC cylindrical mass shell, the
experiment has two independent parameters, namely, the linear mass density,
and the angular velocity of the shell. That means one could shift the
interference fringes by changing the angular velocity. Since the
gravito-electric field from the mass of the shell does not depend on the
angular velocity, a fringe shift will happen despite the fact that the
classical force has not changed. Hence the AB fringe shift in the
gravitational case could not have had a classical origin.

Now from the Maxwell-like equation (\ref{Maxwell-like-eq-3}), and from the
vector identity%
\begin{equation}
\mathbf{\nabla \cdot }\left( \mathbf{\nabla \times h}\right) =0
\end{equation}%
it follows that it is always possible to express the magnetic-like field $%
\mathbf{B}_{g}$ as%
\begin{equation}
\mathbf{B}_{g}=\mathbf{\nabla \times h}  \label{B_g = curl h}
\end{equation}%
for some vector field $\mathbf{h}$. The relationship (\ref{B_g = curl h}) is
formally identical to the relationship in electromagnetism between the
magnetic field $\mathbf{B}$ and the electromagnetic vector potential $%
\mathbf{A}$%
\begin{equation}
\mathbf{B}=\mathbf{\nabla \times A}
\end{equation}%
which follows from the Maxwell equation $\mathbf{\nabla \cdot B}=0$.
Therefore we shall call $\mathbf{h}$ the \textquotedblleft gravitational
vector potential.\textquotedblright

In the gravitational case, just as in the electromagnetic case, the
gravitational vector potential $\mathbf{h}$ possesses the gauge freedom%
\begin{equation}
\mathbf{h\rightarrow h+\nabla }\mu  \label{gauge freedom for h}
\end{equation}%
where $\mu $ can be any arbitrary scalar function of position. This follows
from the vector identity $\mathbf{\nabla \times \nabla }\mu =0$, and is
formally identical to the case of electromagnetism, in which the vector
potential $\mathbf{A}$ possesses the gauge freedom%
\begin{equation}
\mathbf{A\rightarrow A+\nabla }\lambda  \label{gauge freedom for A}
\end{equation}%
where $\lambda $ can be any arbitrary scalar function of position. Again,
this follows from the vector identity $\mathbf{\nabla \times \nabla }\lambda
=0$.

Now the principle of \emph{local} gauge invariance in nonrelativistic
quantum mechanics states that the phase of the time-independent wavefunction 
$\Psi \left( \mathbf{r}\right) $ of any quantum system must always be able
to be \emph{locally} transformed without affecting the physics of the
system. In other words, the transformation \cite{weyl}%
\begin{equation}
\Psi \left( \mathbf{r}\right) \rightarrow \Psi \left( \mathbf{r}\right) \exp
\left( i\phi \left( \mathbf{r}\right) \right)
\label{local gauge transformation}
\end{equation}%
where the phase $\phi \left( \mathbf{r}\right) $ can be any \emph{arbitrary}
real scalar function of position $\mathbf{r}$, can neither change the
properties of the quantum system, nor the physical laws governing the system
and its interactions with its environment. In particular, this \emph{local}
transformation of the phase of the wavefunction cannot change the
probability distribution of the system, since%
\begin{equation}
\left\vert \Psi \left( \mathbf{r}\right) \right\vert ^{2}\rightarrow
\left\vert \Psi \left( \mathbf{r}\right) \exp \left( i\phi \left( \mathbf{r}%
\right) \right) \right\vert ^{2}=\left\vert \Psi \left( \mathbf{r}\right)
\right\vert ^{2}
\end{equation}%
and therefore the Born probability interpretation of the wavefunction is
unaffected by this transformation.

However, \emph{gradients} of $\Psi \left( \mathbf{r}\right) $ will be
changed by the introduction of an \emph{arbitrary} scalar function $\phi
\left( \mathbf{r}\right) $, and therefore will alter the momentum of the
system. If so, one could \emph{arbitrarily} alter the physical laws
governing the system, including altering the conservation of momentum of a
particle in the usual $\exp \left( i\mathbf{p\cdot r}/\hbar \right) $ plane
wave state of an electron within a force-free region of space, where one
knows that $\mathbf{p}$ must be a constant. This obviously cannot be the
case. Therefore the principle of local gauge invariance \emph{necessitates}
the existence of some compensating vector field (or fields), such as the $%
\mathbf{A}$ and $\mathbf{h}$ fields in the DeWitt minimal coupling rule \cite%
{DeWitt}%
\begin{eqnarray}
\frac{\hbar }{i}\mathbf{\nabla } &\rightarrow &\frac{\hbar }{i}\mathbf{\nabla }-q\mathbf{A}-m\mathbf{h} \\
\mathbf{p} &\rightarrow &\mathbf{p}-q\mathbf{A}-m\mathbf{h}  \label{DeWitt minimal coupling}
\end{eqnarray}%
where $\mathbf{p}_{\textrm{op}}=\frac{\hbar }{i}\mathbf{\nabla }$ is the 
%TEXT
momentum operator, $q$ is the charge, and $m$ is the mass of the
nonrelativistic quantum system under consideration. Here $\mathbf{A}$ and $%
\mathbf{h}$ are, respectively, the vector potentials for electromagnetism
and for weak gravitation, which are being viewed here as being the requisite
\textquotedblleft compensating vector fields,\textquotedblright\ whose
existence is \emph{necessitated} by the principle of local gauge invariance.
Since the vector fields $\mathbf{A}$ and $\mathbf{h}$ have the gauge
freedoms $\mathbf{A\left( \mathbf{r}\right) \rightarrow A\left( \mathbf{r}%
\right) +\nabla }\lambda \left( \mathbf{r}\right) $ and $\mathbf{h\left( 
\mathbf{r}\right) \rightarrow h\left( \mathbf{r}\right) +\nabla }\mu \left( 
\mathbf{r}\right) $, these freedoms can then be used to compensate for the
gauge freedom in the transformation $\Psi \left( \mathbf{r}\right)
\rightarrow \Psi \left( \mathbf{r}\right) \exp \left( i\phi \left( \mathbf{r}%
\right) \right) $, in just such a way that the quantum system can once again
satisfy the principle of local gauge invariance.

Thus invoking the DeWitt minimal coupling rule (\ref{DeWitt minimal coupling}%
), we demand that the nonrelativistic Hamiltonian of any quantum system in
the presence of $\mathbf{A}$ and $\mathbf{h}$ fields must always have the
following form \cite{DeWitt}:%
\begin{equation}
H=\frac{1}{2m}\left( \mathbf{p}-q\mathbf{A}-m\mathbf{h}\right) ^{2}+V
\label{complete hamiltonian}
\end{equation}%
where $V$ is the potential energy of the system. Here, in the present
context of the SC quantum systems that we are interested in, such as that of
the SC \textquotedblleft solenoid\textquotedblright\ pictured in Figure 1, $%
q=2e$ is the charge of a Cooper pair, and $m=2m_{e}$ is its mass.

Although one can always \emph{arbitrarily} choose a gauge locally so that
both vector fields $\mathbf{A}$ and $\mathbf{h}$ are set identically equal
to zero at each point exterior to the solenoid, nevertheless the \emph{fluxes%
} interior to the solenoid%
\begin{eqnarray}
\Phi &=&\oint\limits_{C}\mathbf{A}\cdot d\mathbf{l}  \label{magnetic flux} \\
\Phi _{g} &=&\oint_{C}\mathbf{h}\cdot d\mathbf{l}
\label{gravito-magnetic flux}
\end{eqnarray}%
where $C$ is a closed curve enclosing the solenoid, cannot be arbitrarily
set equal to zero, but must instead be gauge-invariant, nonzero, globally 
\emph{measurable} quantities. Hence\ the fluxes $\Phi $ and $\Phi _{g}$ must
be \emph{physical} quantities.

The gravitational Aharonov-Bohm effect depicted in Figure 1 is closely
related to the time holonomy which arises from the off-diagonal
time-space components of the metric tensor $g_{0i}$ \cite{metric conventions}%
. It can be shown \cite{Landau and Lifshitz} that this time holonomy $\Delta
t$\ can be expressed as follows:%
\begin{equation}
\Delta t=-\frac{1}{c}\oint\limits_{C}\frac{g_{0i}}{g_{00}}dx^{i}
\label{exact time holonomy}
\end{equation}%
where $C$ is an arbitrary closed curve in space (such as the one enclosing
the \textquotedblleft solenoid\textquotedblright\ in Figure 1), and $dx^{i}$
is a spatial line element of this closed curve. In light of the time
holonomy given by (\ref{exact time holonomy}), it is impossible in general
relativity to define a global time coordinate for an entire physical system,
such as the topologically nontrivial superconductor in Figure 1.

In the weak field, slow matter approximation, in which%
\begin{equation}
g_{0i}\approx h_{0i}
\end{equation}%
where $h_{0i}$\ are the time-space components of the small-deviation metric $%
h_{\mu \nu }$ from the Minkowski metric $\eta _{\mu \nu }$, and in which the
time-time component of the metric can be approximated by%
\begin{equation}
g_{00}\approx -1
\end{equation}%
it follows that the time holonomy (\ref{exact time holonomy}) becomes
approximately%
\begin{equation}
\Delta t\approx \frac{1}{c}\oint\limits_{C}h_{0i}dx^{i}
\label{approximate time holonomy}
\end{equation}%
For electron waves traveling around a closed curve $C$ enclosing a
\textquotedblleft solenoid\textquotedblright\ such as that in Figure 1, the 
\emph{time} holonomy (\ref{approximate time holonomy}) becomes the \emph{%
phase} holonomy%
\begin{equation}
\Delta \phi =\omega _{\textrm{Compton}}\Delta t\approx \frac{m_{e}c^{2}}{\hbar 
}\frac{1}{c}\oint\limits_{C}h_{0i}dx^{i}\neq 0
\label{phase shift for Compton waves}
\end{equation}%
where $\omega _{\textrm{Compton}}=m_{e}c^{2}/\hbar $ is the Compton frequency 
%TEXT
of the electron \cite{zeilinger}. The phase shift (\ref{phase shift for
Compton waves}) , which is nonvanishing for the \textquotedblleft
solenoid\textquotedblright\ configuration of Figure 1, is the gravitational AB phase shift. It is closely
related to Berry's phase \cite{Berry}, since both phases have a common
origin in non-Euclidean geometry.

Since physically counting the number of fringes in a shift of an
Aharonov-Bohm interference pattern, such that in Figure 1, must yield the
same result for all observers, independent of their state of motion under a 
restricted set of (Galilean)\ coordinate transformations, it is
sufficient for the purposes of this paper to say that the flux $\Phi _{G}$
is a Galilean invariant, and therefore a measurable, \emph{physical}
quantity. Moreover, the closed-path integral of $h_{0i}dx^{i}$ in ({\ref{approximate time holonomy}) is an intrinsic time holonomy which cannot vanish due to an arbitrary gauge transformation.

However, if one were to \emph{arbitrarily} make the global gauge choice%
\begin{equation}
h_{0i}=0\textrm{ everywhere}  \label{TT gauge choice}
\end{equation}%
as is done, for example, in the transverse-traceless gauge, then it follows
that the phase holonomy must vanish identically, i.e.,%
\begin{equation}
\Delta \phi \approx \frac{mc^{2}}{\hbar }\frac{1}{c}\oint%
\limits_{C}h_{0i}dx^{i}=0\textrm{ by setting }h_{0i}=0\textrm{ everywhere}
\label{vanishing of gravitational AB phase}
\end{equation}%
and the gravitational Aharonov-Bohm phase shift predicted for the electron
interference pattern in Figure 1 would disappear.

However, just as in the case of the electromagnetic Aharonov-Bohm effect
where 
\begin{equation}
\Delta \phi =\frac{q}{\hbar }\oint\limits_{C}A_{i}dx^{i}=0\textrm{ by setting }%
A_{i}=0\textrm{ everywhere}  \label{AB phase = 0 by gauge choice}
\end{equation}%
the results (\ref{vanishing of gravitational AB phase}) and (\ref{AB phase =
0 by gauge choice}) are both unphysical whenever the closed curve $C$
encloses either a solenoid\ with a nonvanishing electromagnetic flux $\Phi
\neq 0$, or a \textquotedblleft solenoid\textquotedblright\ with a
nonvanishing Lense-Thirring flux $\Phi _{g}\neq 0$, since both of these
fluxes are gauge-invariant, measurable, physical quantities that cannot be
arbitrarily set equal to zero. Hence the transverse-traceless gauge choice (%
\ref{TT gauge choice}) is unphysical in situations that involve the
gravitational Aharonov-Bohm effect depicted in Figure 1 where $\Phi _{g}\neq
0$, and also, by extension, in time-varying situations that involve $\Phi
_{g}\left( t\right) \neq 0$, for example, in situations where gravitational
radiation is interacting with superconducting systems.

The usual Aharonov-Bohm phase shift follows from DeWitt's minimal coupling
rule (\ref{DeWitt minimal coupling}) when one sets $\Phi _{g}=0$, for then
the phase shift arising from (\ref{DeWitt minimal coupling}) in the
configuration pictured in Figure 1, reduces down to the usual expression%
\begin{equation}
\Delta \phi =\frac{q}{\hbar }\oint\limits_{C}\mathbf{A}\cdot d\mathbf{l}%
\textrm{ }\neq 0
\end{equation}%
and we recover the standard form for the AB phase shift. However, if $\Phi
_{g}\neq 0$, then the total AB phase, upon integration over any closed curve 
$C$ enclosing the \textquotedblleft solenoid,\textquotedblright\ becomes%
\begin{equation}
\Delta \phi _{\textrm{tot}}=\frac{q}{\hbar }\oint\limits_{C}\mathbf{A}\cdot d%
\mathbf{l}\textrm{ }+\frac{m}{\hbar }\oint_{C}\mathbf{h}\cdot d\mathbf{l=}%
\frac{q\Phi }{\hbar }+\frac{m\Phi _{g}}{\hbar }  \label{total AB phase}
\end{equation}%
The vector potential $\mathbf{h}$ in (\ref{total AB phase}) can arise either
from a Lense-Thirring field, or from rotations of a quantum system, such as
from a rotating SC ring. Since experiments with rotating SC systems are much
easier to perform than experiments involving Lense-Thirring fields, we shall
confine our attention for now to these much easier experiments.

One can check experimentally the expression given by (\ref{total AB phase})
for a rotating SC ring, in which case the single-valuedness of the
macroscopic wavefunction of the Cooper pairs demands that $\Delta \phi _{%
\textrm{tot}}=2\pi n$, where $n$ is an integer. It follows from this that a
magnetic field must be generated by rotating a superconducting ring, i.e.,
that a \textquotedblleft London moment\textquotedblright\ must accompany
this rotation. Precision measurements of the London moment of a rotating SC
ring thus provide a test for the correctness of the expression for the total
AB phase\ in (\ref{total AB phase}). Cabrera and co-workers \cite{cabrera}
performed these measurements to 100 parts per million. Thus the formula in (%
\ref{total AB phase}) for the total AB phase has been experimentally
verified. In this way, the expression in (\ref{DeWitt minimal coupling}) for
the DeWitt minimal coupling rule has been experimentally tested to high
precision.

So far we have been considering only the case of stationary,
time-independent, charge and mass currents, such as those in a SC magnet, or
in a steadily rotating SC ring. These quantum currents can be the quantum
mechanical sources of time-independent $\mathbf{A}$ and $\mathbf{h}$ fields that give rise to the AB effect.

\section{The dynamical Casimir effect via parametric oscillations}

In this section, we begin by discussing the case when the vector potentials $%
\mathbf{A}$ and $\mathbf{h}$ are time dependent, i.e. $\mathbf{A}(\mathbf{r}%
,t)$ and $\mathbf{h}(\mathbf{r},t)$. From this we will propose a version of
the dynamical Casimir effect for the electromagnetic vector potential $%
\mathbf{A}$, in which photons are \textquotedblleft
pumped\textquotedblright\ out of the vacuum via parametric oscillations of a
SC membrane.

It is natural to extend the time-independent Hamiltonian (\ref{complete
hamiltonian}) to the following time-dependent one \cite{DW-limits}: 
\begin{equation}
H=\frac{1}{2m}\left( \mathbf{p}-q\mathbf{A}\left( \mathbf{r},t\right) -m%
\mathbf{h}\left( \mathbf{r},t\right) \right) ^{2}+V
\label{time-dependent complete Hamiltonian}
\end{equation}%
in which the fields $\mathbf{A}\left( \mathbf{r},t\right) $ and $\mathbf{h}%
\left( \mathbf{r},t\right) $ are to be first treated as classical fields,
but the matter (e.g., the vibrating SC wire in Figure 2) is to be treated
quantum mechanically, in the so-called \textquotedblleft semi-classical
approximation.\textquotedblright\ 

Expanding the square in (\ref{time-dependent complete Hamiltonian}), one
obtains the following interaction Hamiltonian terms:%
\begin{equation}
H_{\mathbf{p\cdot A}}=-\frac{q}{m}\mathbf{p\cdot A}\left( \mathbf{r},t\right)
\end{equation}%
which leads to the interaction of the quantum system with electromagnetic
(EM) radiation, such as in the stimulated emission and absorption of EM
waves by the quantum matter, and%
\begin{equation}
H_{\mathbf{p\cdot h}}=-\mathbf{p\cdot h}\left( \mathbf{r},t\right)
\end{equation}%
which leads to the interaction of the quantum system with gravitational (GR)
radiation, such as in the stimulated emission and absorption of GR waves by
the quantum matter, and%
\begin{equation}
H_{\mathbf{A\cdot h}}=+q\mathbf{A}\left( \mathbf{r},t\right) \mathbf{\cdot h}%
\left( \mathbf{r},t\right)
\end{equation}%
which leads to the interaction between EM and GR radiation fields mediated
by the quantum system, such as in the transduction of EM waves into GR waves
mediated by the quantum matter \cite{Chiao-Townes-volume}, and%
\begin{equation}
H_{\mathbf{A\cdot A}}=+\frac{q^{2}}{2m}\mathbf{A}\left( \mathbf{r},t\right) 
\mathbf{\cdot A}\left( \mathbf{r},t\right)  \label{H_A.A}
\end{equation}%
which leads to Landau-diamagnetism type of interactions of the quantum
system with EM radiation, such as in the parametric amplification of EM
waves by a strongly driven, i.e., \textquotedblleft
pumped,\textquotedblright\ quantum system (see, for example, Figures 2 and
4), and%
\begin{equation}
H_{\mathbf{h\cdot h}}=+\frac{m}{2}\mathbf{h}\left( \mathbf{r},t\right) 
\mathbf{\cdot h}\left( \mathbf{r},t\right)  \label{H_h.h}
\end{equation}%
which leads to \emph{gravitational} Landau-diamagnetism type of interactions
of the quantum system with GR radiation, such as in the parametric
amplification of GR waves, again by a strongly driven, i.e.,
\textquotedblleft pumped,\textquotedblright\ quantum system (again, see, for
example, Figures 2 and 4).

All of the above interaction terms will be treated as small perturbations of
the unperturbed Hamiltonian%
\begin{equation}
H_{0}=\frac{\mathbf{p}^{2}}{2m}+V
\end{equation}%
and can thus be treated using standard perturbation theory.

At a fully quantum mechanical level of description, both the matter and the
radiation fields $\mathbf{A}$ and $\mathbf{h}$ would have to be quantized.
The radiation fields could be quantized by invoking the commutation relations%
\begin{equation}
\lbrack a,a^{\dag }]=1  \label{[a,a dagger] = 1}
\end{equation}%
\begin{equation}
\lbrack b,b^{\dag }]=1  \label{[b,b dagger] = 1}
\end{equation}%
where $a$ and $a^{\dag }$ are, respectively, the annihilation and creation
operators for a quantum of a single mode of the EM radiation field, and
where $b$ and $b^{\dag }$ are, respectively, the annihilation and creation
operators for a quantum of a single mode of the GR radiation field.

For now, let us focus solely on the interactions of quantized EM radiation
with matter. The second quantized form for the EM vector potential operator $%
A_{\textrm{op}}\left( \mathbf{r}\right) $, when summed over all the modes of a
cavity enumerated by the index $\kappa $, is \cite{garrison&chiao} 
\begin{equation}
A_{\textrm{op}}\left( \mathbf{r}\right) =\sum\limits_{\kappa }\sqrt{\frac{%
\hbar }{2\varepsilon _{0}\omega _{\kappa }}}\left( a_{\kappa }+a_{\kappa
}^{\dag }\right) \mathcal{E}_{\kappa }\left( \mathbf{r}\right)
\end{equation}%
where $\omega _{\kappa }$ is the frequency of mode $\kappa $, and $\mathcal{E%
}_{\kappa }\left( \mathbf{r}\right) $ is one Cartesian component of the
classical electric field distribution associated with this mode. Therefore,
when there is a single dominant mode in the problem, the vector potential
operator simplifies to the expression 
\begin{equation}
A_{\textrm{op}}\left( \mathbf{r}\right) \propto \left( a+a^{\dag }\right)
\end{equation}%
where the mode index $\kappa $ and the proportionality constant have been
suppressed.

The expansion of the square in the interaction Hamiltonian $H_{\mathbf{%
A\cdot A}}\propto \left( a+a^{\dag }\right) ^{2}$ in (\ref{H_A.A}) will
therefore contain the term \cite{garrison2}%
\begin{equation}
K_{\textrm{op}}\propto a^{\dag }a^{\dag }+\textrm{hermitian adjoint}
\label{generator of squeezed state}
\end{equation}%
The $a^{\dag }a^{\dag }$ term of the operator $K_{\textrm{op}}$ corresponds to
the process of photon \emph{pair creation} in the parametric amplification
arising from the pumping action of some strong \textquotedblleft
pump\textquotedblright\ wave upon a quantum system. It can be shown that (%
\ref{generator of squeezed state}) has the form of an infinitesimal
generator of a squeezed state of light \cite{garrison2}.

\begin{figure}[h]
%\sidecaption[t]
\centering
\includegraphics[angle=0,width=.6%
\textwidth]{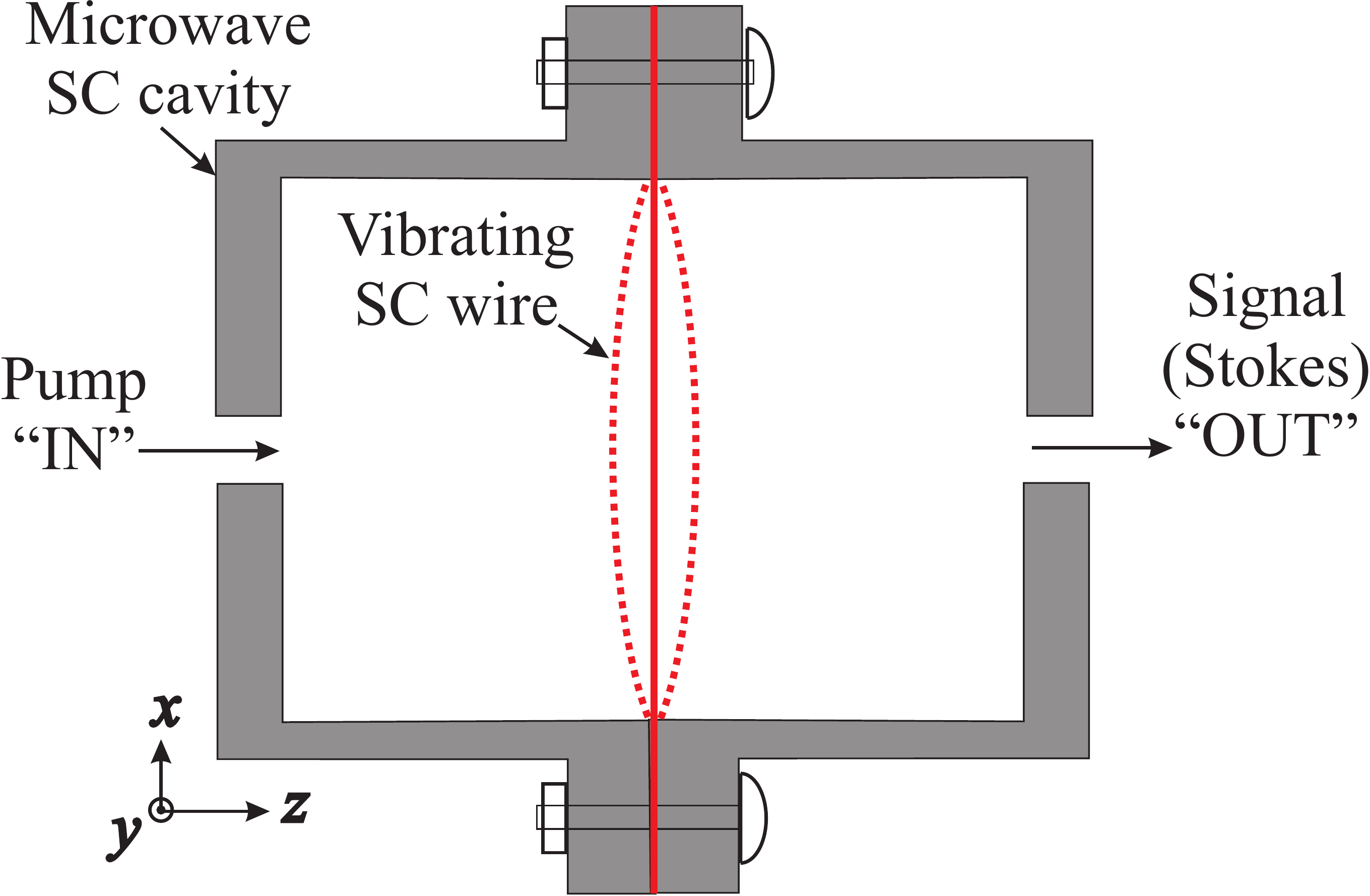} %\label{fig:REFERENCE}
\caption{(Color online) A parametric amplifier or oscillator, whose active
element is the vibrating SC\ wire (in red) placed in the middle of a
microwave SC cavity (in grey). The moving wire can be viewed as if it were
an oscillating \textquotedblleft semipermeable membrane,\textquotedblright\
which does work upon some \textquotedblleft seed\textquotedblright\
radiation initially present in the cavity, thus amplifying this radiation in
a reciprocating, piston-like action. Photons incident upon this moving
\textquotedblleft membrane\textquotedblright\ experience a Doppler shift
that changes their energy. Thus when pump microwaves enter through the left
hole, an amplified signal (Stokes)\ wave will exit through the right hole.}
\end{figure}

%\FRAME{fhFU}{4.4348in%
%}{2.9023in}{0pt}{\Qcb{A parametric amplifier or oscillator, whose active
%element is the vibrating SC\ wire (in red) placed in the middle of a
%microwave SC cavity (in grey). The moving wire can be viewed as if it were
%an oscillating \textquotedblleft semipermeable membrane,\textquotedblright\
%which does work upon some \textquotedblleft seed\textquotedblright\
%radiation initially present in the cavity, thus amplifying this radiation in
%a reciprocating, piston-like action. Photons incident upon this moving
%\textquotedblleft membrane\textquotedblright\ experience a Doppler shift
%that changes their energy. Thus when pump microwaves enter through the left
%hole, an amplified signal (Stokes)\ wave will exit through the right hole.}}{%
%}{double-cavity-with-vibrating-wire-a.eps}{\special{language "Scientific
%Word";type "GRAPHIC";maintain-aspect-ratio TRUE;display "USEDEF";valid_file
%"F";width 4.4348in;height 2.9023in;depth 0pt;original-width
%10.8776in;original-height 7.1027in;cropleft "0";croptop "1";cropright
%"1";cropbottom "0";filename
%'double-cavity-with-vibrating-wire-a.eps';file-properties "NPEU";}}

Instead of enumerating all the possible resulting second quantized forms of
the above interaction Hamiltonian terms, let us just focus on one such term,
namely, $H_{\mathbf{A\cdot A}}$ in (\ref{H_A.A}), which is associated with
parametric amplification, such as that in the setup depicted in Figure 2.
This Figure represents an \textquotedblleft
opto-mechanical\textquotedblright\ parametric amplifier, which becomes,
above threshold, a parametric oscillator, whose active element is the
central vibrating SC wire (indicated in red), placed across the middle of an
extremely high-$Q$ SC microwave cavity. Here, instead of using optical
cavities, as is usual in ongoing opto-mechanical experiments \cite%
{aspelmeyer}, we shall be using SC microwave cavities. The reason for this
is that the quality factor for SC microwave cavities has already been
demonstrated by Haroche and co-workers \cite{kuhr07} to be on the order of%
\begin{equation}
Q\sim 10^{10}
\end{equation}%
which can be much higher than that of typical optical cavities.

The motion of the SC wire in the middle of the microwave cavity will
modulate the \textquotedblleft pump\textquotedblright\ microwaves coming
through the \textquotedblleft IN\textquotedblright\ port so as to produce
radiation at new sideband frequencies via the Doppler effect. The
\textquotedblleft seed\textquotedblright\ radiation initially in one of
these sidebands, namely, the first \textquotedblleft
Stokes\textquotedblright\ sideband, can then become the exponentially
amplified. Macroscopic, easily detectable radiation in the form of a strong
Stokes wave emitted by the parametric amplifier can then leave the cavity
through the \textquotedblleft OUT\textquotedblright\ port. This kind of
strong Stokes emission would be similar to the Stokes emission observed in
the stimulated Raman effect in nonlinear optics \cite{Boyd}.

However, here we shall first analyze classically the parametric
amplification process in Figure 2, in order to answer the following
questions: What is the threshold for parametric oscillation in Figure 2? Is
this experiment feasible to perform? The key concept that we shall use in
this classical analysis is that of the \emph{work done} by the moving wire,
viewed as if the wire were a moving \textquotedblleft
piston\textquotedblright\ acting on some \textquotedblleft
seed\textquotedblright\ radiation initially present in the cavity. This
\textquotedblleft piston-like\textquotedblright\ action of the moving wire
can also be viewed as if the wire were a partially reflecting
\textquotedblleft moving mirror,\textquotedblright\ and will lead to the 
\emph{exponential} amplification of the \textquotedblleft
seed\textquotedblright\ radiation at the Stokes frequency above the
threshold of parametric oscillation.

The reason for using a vibrating wire instead of a vibrating membrane is
that the wire is one dimensional, whereas the membrane is two dimensional.
The mass of a thin wire can be made much smaller than the mass of a thin
membrane, and therefore a wire can be driven more easily into motion.

However, here we shall model the vibrating wire in Figure 2 as if the wire
were a vibrating \textquotedblleft semi-permeable
membrane,\textquotedblright\ for which the pressure acting on the membrane
can be converted into an easily calculable force. The justification for this
\textquotedblleft membrane\textquotedblright\ model is that the scattering
cross section of a thin conducting wire, when placed symmetrically across
the mouth of a waveguide, can be comparable in size to the cross-sectional
area of the waveguide, because the wire tends to \textquotedblleft short
out\textquotedblright\ the electric field of the TE mode of a waveguide.
Thus the reflection coefficient of the wire placed across the middle of a
cavity (as in Figure 2), can be made quite high. There results a splitting
of a microwave cavity mode into a spectral doublet as illustrated
in Figure 3 \cite{FQMT11}\cite{2 interacting microwave resonances}, due to
the presence of the central wire. The splitting frequency of the doublet can
typically be on the order of 1 GHz for a microwave mode frequency
of around 10 GHz.

\begin{figure}[h]
%\sidecaption[t]
\centering
\includegraphics[angle=0,width=.7\textwidth]{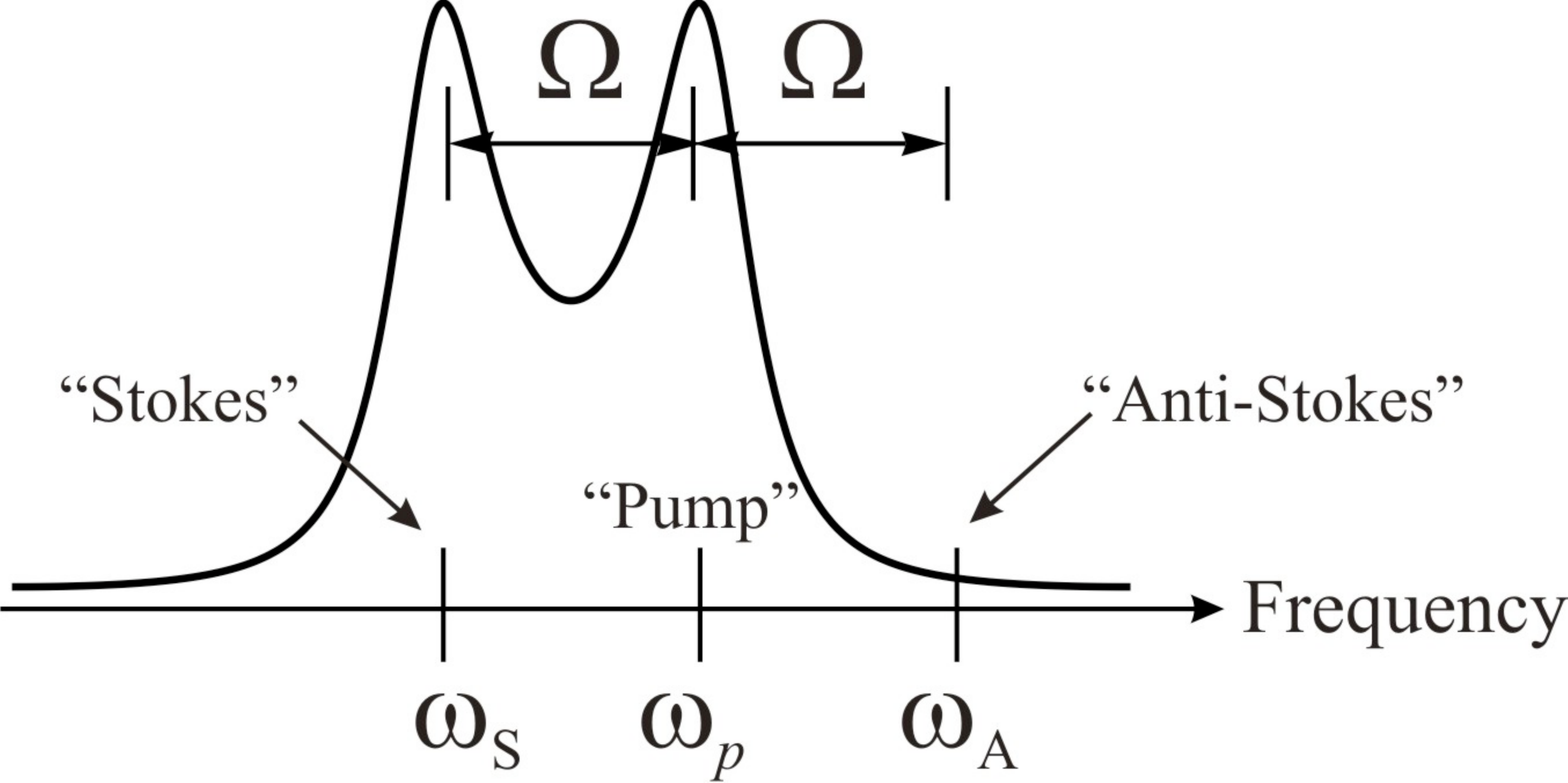} 
%\label{fig:REFERENCE}
\caption{The excitation spectrum of a microwave cavity with a wire placed at
its center (see Figure 2). The spectrum consists of a doublet of resonances
at the \textquotedblleft signal\textquotedblright\ mode at the
\textquotedblleft Stokes\textquotedblright\ frequency $\protect\omega _{%
\textrm{S}}$ and the \textquotedblleft pump\textquotedblright\ mode at the
\textquotedblleft pump\textquotedblright\ frequency $\protect\omega _{p}$.
The difference between \textquotedblleft pump\textquotedblright\ and
\textquotedblleft Stokes\textquotedblright\ frequencies\ is resonant with
the frequency $\Omega \ $of the vibrating wire\ $\left( \textrm{{i.e., }}%
\protect\omega _{p}-\protect\omega _{\textrm{S}}=\Omega \right) $. The
\textquotedblleft anti-Stokes\textquotedblright\ frequency\ $\protect\omega %
_{\textrm{A}}$ is off resonance with respect to the doublet, and hence is
suppressed. (Cf. the stimulated Raman effect in \protect\cite{Boyd}).}
\end{figure}

%\FRAME{ftbpFU}{%
%4.5809in}{2.2935in}{0pt}{\Qcb{The excitation spectrum of a microwave cavity
%with a wire placed at its center (see Figure 2). The spectrum consists of a
%doublet of resonances at the \textquotedblleft signal\textquotedblright\
%mode at the \textquotedblleft Stokes\textquotedblright\ frequency $\protect%
%\omega _{\textrm{S}}$ and the \textquotedblleft pump\textquotedblright\ mode
%at the \textquotedblleft pump\textquotedblright\ frequency $\protect\omega %
%_{p}$. The difference between \textquotedblleft pump\textquotedblright\ and
%\textquotedblleft Stokes\textquotedblright\ frequencies\ is resonant with
%the frequency $\Omega \ $of the vibrating wire\ $\left( \textrm{{i.e., }}%
%\protect\omega _{p}-\protect\omega _{\textrm{S}}=\Omega \right) $. The
%\textquotedblleft anti-Stokes\textquotedblright\ frequency\ $\protect\omega %
%_{\textrm{A}}$ is off resonance with respect to the doublet, and hence is
%suppressed. (Cf. the stimulated Raman effect in \protect\cite{Boyd}).}}{}{%
%dspectrum-a.eps}{\special{language "Scientific Word";type
%"GRAPHIC";maintain-aspect-ratio TRUE;display "USEDEF";valid_file "F";width
%4.5809in;height 2.2935in;depth 0pt;original-width 4.529in;original-height
%2.2537in;cropleft "0";croptop "1";cropright "1";cropbottom "0";filename
%'Dspectrum-a.eps';file-properties "NPEU";}}

We have observed the splitting of a microwave cavity resonance into a
spectral doublet. An RF cylindrical copper cavity of length $L= 1.284"$ and
diameter $D=1.02"$ with two parallel conducting end plates was constructed
to support a TE112 mode at 11.42 GHz \cite{Bell paper}. (The mode indices $%
l,m,n$ are chosen to correspond to the number of half wavelengths along their
respective axes; angular, radial, and axial respectively.) The input
coupler, a short straight wire placed perpendicular to the axial direction,
was placed on the cylinder at approximately a quarter wavelength from one
end plate. The output coupler is a small loop placed at the other end plate
of the cavity.

An off-diagonal scattering matrix element S21 transmission measurement, preformed with a HP 8720C network analyzer,
shows the resonance of the TE112 mode at approximately 11.50 GHz; see figure %
\ref{TE112}(a). A splitting is observed by placing a 22AWG copper wire at
the midpoint of the cavity, perpendicular to the axial direction and
parallel to the input coupler. The splitting is approximately 400 MHz; see
figure \ref{TE112}(b).

\begin{figure}[h]
%\sidecaption[t]
% For example, with the graphicx style use
\centering
\includegraphics[scale=.3]{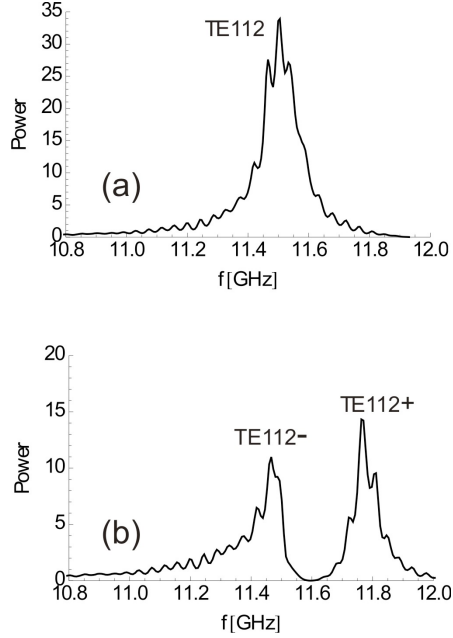} %
% If no graphics program available, insert a blank space i.e. use
%\picplace{5cm}{2cm} % Give the correct figure height and width in cm
%
% Give a unique label
\caption{Splitting of the TE112 mode in cavity with a bisecting copper wire
at its midpoint perpendicular to the axial direction and parallel to the
input coupler. The splitting is on the order of 400 MHz. S21 transmission
measurements (a) of a copper cavity with TE112 resonant frequency at 11.5
GHz and (b) splitting due to the placement of copper wire placed at the
center. The vertical axes use the same arbitrary power reference in the
conversion from logarithmic to linear scale.}
\label{TE112}
\end{figure}

In Figure 3, the pump frequency $\omega _{p}$ of the parametric amplifier is
assumed to be tuned to coincide with the upper member of the spectral
doublet, and the signal frequency $\omega _{\textrm{S}}$ is assumed to be
tuned to coincide with the lower member of this doublet, which we shall call
the \textquotedblleft Stokes frequency,\textquotedblright\ in analogy with
the stimulated Raman effect \cite{Boyd}. The idler frequency $\omega _{i}$,
i.e., the frequency of the mechanical motion of the central wire in Figure
2, is the beat frequency $\Omega =\omega _{p}-\omega _{\textrm{S}}$ between
the pump and signal frequencies. Note that the parasitic, Doppler upshifted
\textquotedblleft anti-Stokes frequency\textquotedblright\ $\omega _{\textrm{A}%
}$\ is automatically suppressed by this spectral doublet.

To calculate the force acting on the central membrane (as a model of the
force acting on the central wire) in the middle of the microwave cavity of
Figure 2, we begin from the Maxwell stress tensor \cite{griffiths}\cite%
{jackson}%
\begin{equation}
T_{ij}=\varepsilon _{0}\left( E_{i}E_{j}-\frac{1}{2}\delta _{ij}E^{2}\right)
+\frac{1}{\mu _{0}}\left( B_{i}B_{j}-\frac{1}{2}\delta _{ij}B^{2}\right)
\label{maxwell stress tensor}
\end{equation}

Now starting from the electromagnetic force exerted on charges and currents
given by the Lorentz force law%
\begin{equation}
\mathbf{F}=q\left( \mathbf{E+v\times B}\right)
\end{equation}%
it can be shown that there results the following relationship between the
force $\mathbf{F}$ and the total Maxwell stress tensor $T_{ij}$ and the
Poynting vector $\mathbf{S}$\ \cite{griffiths}\cite{jackson}:%
\begin{equation}
\left( \mathbf{F}\right) _{i}=\oint\limits_{S\left( V\right) }T_{ij}\cdot
\left( d\mathbf{a}\right) ^{j}-\varepsilon _{0}\mu _{0}\frac{d}{dt}%
\int\limits_{V}\left( \mathbf{S}\right) _{i}dV\textrm{ (where }i=1,2,3\textrm{)}
\label{surface and volume integrals}
\end{equation}%
where $T_{ij}$ is the stress tensor evaluated at $d\mathbf{a}$, an
infinitesimal area element of an arbitrary surface $S\left( V\right) $ which
encloses the volume $V$, $dV$ is an infinitesimal volume element of the
matter inside the volume $V$ enclosed by the surface $S\left( V\right) $,
and $\mathbf{S=E\times H}$ is the Poynting vector evaluated at $dV$ inside $%
V $.

Now the tangential electric field must vanish at the boundary of any
conductor. Hence, for all transverse electric modes of the microwave cavity
pictured in Figure 2, if one chooses the surface $S\left( V\right) $ to be
that of a small pillbox straddling a patch of the surface of an equivalent
SC membrane,\ the contribution to the force (\ref{surface and volume
integrals}) from the Poynting vector term evaluated at the pillbox enclosing
the patch of the surface, must vanish. In the case of transverse magnetic
modes of the cavity, the Poynting vector $\mathbf{S=E\times H}$ does not
vanish at the surface, since there will be a longitudinal component of the
electric field at the surface along with a tangential component of the
magnetic field. Hence there will arise a \emph{tangential} component of $%
\mathbf{S}$ at the surface of the SC membrane, but this $\mathbf{S}$ cannot
contribute to any \emph{normal} force acting on the conducting surface.

Therefore the only contribution to the normal force acting on the equivalent
SC membrane\ arises solely from the Maxwell stress tensor term of (\ref%
{surface and volume integrals}), which, for the case of transverse electric
modes evaluated at the membrane, reduces down to 
\begin{equation}
\left( T_{ij}\right) =\frac{1}{2\mu _{0}}\left( 
\begin{array}{ccc}
-B_{y}^{2} & 0 & 0 \\ 
0 & +B_{y}^{2} & 0 \\ 
0 & 0 & -B_{y}^{2}%
\end{array}%
\right)  \label{Tij for magnetic case}
\end{equation}%
because the magnetic field of the transverse electric mode (e.g., the TE112
mode), whose electric field is pointing in the $x$ direction (in the
Cartesian coordinate system of Figure 2), will be pointing in the $y$
direction at the surface of the equivalent membrane \cite{Larmor}.

If one therefore replaces the SC wire by an equivalent SC membrane,\ then
the diagonal terms of (\ref{Tij for magnetic case}) can be interpreted as a
\textquotedblleft field pressure\textquotedblright\ acting on the membrane
with a maximum amplitude of%
\begin{equation}
P_{\max }=\frac{1}{2\mu _{0}}\left( B^{2}\right) _{\max }=\left(
u_{B}\right) _{\max }  \label{magnetic pressure}
\end{equation}%
where $\left( B^{2}\right) _{\max }$is the maximum of the square of the
magnetic field, and $\left( u_{B}\right) _{\max }$ is the maximum magnetic
energy density, evaluated at the surface of the membrane.

Because the stress tensor depends \emph{quadratically} on the field, there
results a pressure being exerted on the membrane at a beat note frequency
due to the beating between the fields at pump frequency $\omega _{p}$ and
the fields at the Stokes frequency $\omega _{\textrm{S}}$ in the spectral
doublet of Figure 4. This beat note can drive the membrane (or the wire) at
the beat frequency $\Omega =\omega _{p}-\omega _{\textrm{S}}$, i.e., at the
splitting of the upper and lower members of the doublet. The force in the $z$
direction acting on the membrane (or wire) at the beat frequency $\Omega $
will therefore have the form%
\begin{eqnarray}
F_{\Omega } &=&\frac{1}{\mu _{0}}\mathcal{B}_{p}\mathcal{B}_{\textrm{S}}^{\ast
}\exp \left( -i\left( \omega _{p}-\omega _{s}\right) t\right) \cdot \mathcal{%
A}_{\textrm{eff}}+\textrm{c.c.}  \nonumber \\
&\propto &\exp \left( -i\Omega t\right) +\textrm{c.c.}
\label{force at beat frequency}
\end{eqnarray}%
where $\mathcal{A}_{\textrm{eff}}$ is an effective area of the membrane (or,
equivalently, the effective scattering cross-section of the wire), and where%
\begin{equation}
B_{p}=\mathcal{B}_{p}\exp \left( -i\omega _{p}t\right) +\textrm{c.c.}
\label{pump wave}
\end{equation}%
is the pump waveform, with $\mathcal{B}_{p}$\ being the complex amplitude
for the pump magnetic field waveform, and where%
\begin{equation}
B_{\textrm{S}}=\mathcal{B}_{\textrm{S}}\exp \left( -i\omega _{\textrm{S}}t\right) +%
\textrm{c.c.}  \label{Stokes wave}
\end{equation}%
is the Stokes waveform of some small amount of \textquotedblleft
seed\textquotedblright\ radiation already present inside the cavity, with $%
\mathcal{B}_{\textrm{S}}$\ being the complex amplitude for the Stokes magnetic
field waveform. (Note that such \textquotedblleft seed\textquotedblright\
radiation could in principle be vacuum fluctuations of the EM field inside
the cavity.) In the expression (\ref{force at beat frequency}) for the force 
$F_{\Omega }$ at the beat frequency $\Omega $, we have assumed that the pump
wave (\ref{pump wave}) is always much stronger that the \textquotedblleft
seed\textquotedblright\ Stokes wave (\ref{Stokes wave}), i.e., $\left\vert 
\mathcal{B}_{p}\right\vert >>\left\vert \mathcal{B}_{\textrm{S}}\right\vert $.

Since the driving force $F_{\Omega }$ at the beat frequency $\Omega $ can be
made resonant with the acoustical resonance frequency of the membrane, we
shall model the resulting motion of the membrane as that of a simple
harmonic oscillator with a resonance frequency of $\Omega $. Using Newton's
equation of motion for a damped simple harmonic oscillator moving in the $z$
direction, viz.,%
\begin{equation}
m\left( \ddot{z}+\gamma \dot{z}+\Omega ^{2}z\right) =F_{\Omega }
\label{Newton's EOM for free mass}
\end{equation}%
where $m$ is the mass of the membrane, and $\gamma $ is its damping
coefficient, and using an Ansatz of the form%
\begin{equation}
z=z_{\max }\exp \left( -i\Omega t\right) +\textrm{c.c.}
\end{equation}%
for the displacement of the membrane in the $z$ direction, one finds that
its maximum, on-resonancce complex displacement amplitude is%
\begin{equation}
z_{\max }=i\frac{\mathcal{B}_{p}\mathcal{B}_{\textrm{S}}^{\ast }}{\mu
_{0}m\gamma \Omega }\mathcal{A}_{\textrm{eff}}
\end{equation}%
The velocity of the membrane in the $z$ direction will then have the form%
\begin{equation}
v=v_{\max }\exp \left( -i\Omega t\right) +\textrm{c.c.}
\end{equation}%
where the complex velocity amplitude $v_{\max }=-i\Omega z_{\max }$ becomes,
on resonance,%
\begin{equation}
v_{\max }=\frac{\mathcal{B}_{p}\mathcal{B}_{\textrm{S}}^{\ast }}{\mu
_{0}m\gamma }\mathcal{A}_{\textrm{eff}}
\end{equation}%
There results a Doppler effect arising from the velocity of the membrane
moving in the $z$ direction in Figure 2, giving rise to upper and lower
Doppler sidebands. However, only the lower Doppler sideband will be excited,
since only the lower sideband will be resonant with lower member of the
doublet in Figure 4.

The maximum, on-resonance time-averaged power $\left\langle \mathtt{P}%
\right\rangle $ being delivered from the pump wave into the simple-harmonic
membrane motion at the beat frequency $\Omega $, and therefore into the
lower Dopper sideband, i.e., into the Stokes wave (\ref{Stokes wave}), is%
\begin{equation}
\left\langle \mathtt{P}_{\max }\right\rangle =\left\langle F_{\Omega }\cdot
v\right\rangle _{\max }=2\left\vert \mathcal{B}_{p}\right\vert
^{2}\left\vert \mathcal{B}_{\textrm{S}}\right\vert ^{2}\frac{1}{\mu
_{0}^{2}m\gamma }\mathcal{A}_{\textrm{eff}}^{2}  \label{power <F.v>}
\end{equation}

Now invoking the conservation of energy, we find that, if we for the moment
neglect all losses, the Stokes wave will be amplified by this power
transfer, such that the power $\left\langle \mathtt{P}_{\max }\right\rangle $
being tranferred into the Stokes wave must equal the rate of growth of the
time-averaged Stokes energy inside the cavity%
\begin{equation}
\left\langle U_{\textrm{S}}\right\rangle =\frac{1}{\mu _{0}}\left\vert 
\mathcal{B}_{\textrm{S}}\right\vert ^{2}V_{\textrm{eff}}=\frac{1}{\mu _{0}}%
\left\vert \mathcal{B}_{\textrm{S}}\right\vert ^{2}\mathcal{A}_{\textrm{eff}}%
\mathcal{L}_{\textrm{eff}}  \label{time-average Stokes wave energy in cavity}
\end{equation}%
where $\mathcal{L}_{\textrm{eff}}$ is an effective length of the cavity (i.e., 
$V_{\textrm{eff}}=\mathcal{A}_{\textrm{eff}}\mathcal{L}_{\textrm{eff}}$ is an
effective volume of the cavity). In other words, from energy conservation it
follows that%
\begin{equation}
\left\langle \mathtt{P}_{\max }\right\rangle =\frac{d}{dt}\left\langle U_{%
\textrm{S}}\right\rangle
\end{equation}%
Substituting in from (\ref{power <F.v>}) and (\ref{time-average Stokes wave
energy in cavity}), one infers that%
\begin{equation}
2\left\vert \mathcal{B}_{p}\right\vert ^{2}\left\vert \mathcal{B}_{\textrm{S}%
}\right\vert ^{2}\frac{1}{\mu _{0}^{2}m\gamma }\mathcal{A}_{\textrm{eff}}^{2}=%
\frac{d}{dt}\left( \frac{1}{\mu _{0}}\left\vert \mathcal{B}_{\textrm{S}%
}\right\vert ^{2}\mathcal{A}_{\textrm{eff}}\mathcal{L}_{\textrm{eff}}\right)
=\kappa _{\textrm{S}}\left( \frac{1}{\mu _{0}}\left\vert \mathcal{B}_{\textrm{S}%
}\right\vert ^{2}\mathcal{A}_{\textrm{eff}}\mathcal{L}_{\textrm{eff}}\right)
\label{ode with kappa_gain}
\end{equation}%
where $\kappa _{\textrm{S}}$ is the exponential gain coefficient for
parametric amplification of the Stokes wave. Thus we arrive at an
exponential-growth ODE for the energy stored $U_{\textrm{S}}$\ in the cavity
at the Stokes frequency%
\begin{equation}
\frac{d}{dt}\left\langle U_{\textrm{S}}\right\rangle =\kappa _{\textrm{S}%
}\left\langle U_{\textrm{S}}\right\rangle
\end{equation}%
Solving for the gain coefficient $\kappa _{\textrm{S}}$ from (\ref{ode with
kappa_gain}), we conclude that%
\begin{equation}
\kappa _{\textrm{S}}=\frac{2\left\vert \mathcal{B}_{p}\right\vert ^{2}\mathcal{%
A}_{\textrm{eff}}}{\mu _{0}m\gamma \mathcal{L}_{\textrm{eff}}}
\label{gain coefficient for Stokes}
\end{equation}%
Thus the gain of the Stokes wave is directly proportional to the pump power
stored in the cavity, just like in the stimulated Raman effect \cite{Boyd}.

Like in a laser, the threshold for oscillation \cite{braginsky01}\cite%
{threshold paramp}\ occurs when 
\begin{equation}
\textrm{Gain}=\textrm{Loss}  \label{gain=loss}
\end{equation}%
Above threshold, i.e., when the gain exceeds the loss, macroscopic amounts
of coherent radiation can be produced inside the cavity by the \emph{%
exponential} amplification of the \textquotedblleft seed\textquotedblright\
radiation, i.e., of vacuum fluctuations. Thus even if one were to start off
only with vacuum fluctuations as the \textquotedblleft
seed,\textquotedblright\ one can produce macroscopic amounts of coherent
radiation by the stimulated emission of radiation, just like in a laser. In
principle, this should also apply to GR radiation, as well as to EM
radiation. Thus \emph{generators} of microwave frequency GR radiation should 
\emph{in principle} be possible to construct, as well as amplifiers and
detectors for this kind of radiation. The only remaining question is whether
such devices are feasible \emph{in practice}.

The loss of the Stokes wave (i.e., the \textquotedblleft
signal\textquotedblright ) from the cavity depicted in Figure 2 can result
from emission of radiation through an outcoupling hole into the environment,
or from remnant ohmic losses in the components of the cavity. We shall call
the resulting quality factor of the cavity at the Stokes frequency under
working conditions (i.e., including internal losses and the outcoupling into
the environment) the \textquotedblleft loaded $Q$\textquotedblright , which
is defined as follows:%
\begin{equation}
Q_{\textrm{S}}=\omega _{\textrm{S}}\tau _{\textrm{S}}
\end{equation}%
where \textquotedblleft S\textquotedblright\ stands for \textquotedblleft
Stokes wave\textquotedblright\ whose frequency is $\omega _{\textrm{S}}$, and
whose stored energy inside the cavity decays away with a time scale $\tau _{%
\textrm{S}}$ (the so-called \textquotedblleft cavity ring-down
time\textquotedblright ) after the pump has been turned off.

Moreover, the loss coefficient $\gamma $ in the motion of the simple
harmonic oscillator leads a decay time $\tau _{\Omega }=1/\gamma $ for the
energy stored in the oscillator. This leads to a mechanical oscillator
quality factor $Q_{\Omega }$, which is defined as follows:%
\begin{equation}
Q_{\Omega }=\Omega \tau _{\Omega }
\end{equation}

It follows from the gain-equals-loss condition (\ref{gain=loss})\ that at
threshold%
\begin{equation}
\left( \kappa _{\textrm{S}}\right) _{\textrm{threshold}}=\frac{2\left\vert 
\mathcal{B}_{p}\right\vert _{\textrm{threshold}}^{2}\mathcal{A}_{\textrm{eff}}}{%
\mu _{0}m\gamma \mathcal{L}_{\textrm{eff}}}=\frac{2\left\vert \mathcal{B}%
_{p}\right\vert _{\textrm{threshold}}^{2}Q_{\Omega }\mathcal{A}_{\textrm{eff}}}{%
\mu _{0}m\Omega \mathcal{L}_{\textrm{eff}}}=\frac{1}{\tau _{\textrm{S}}}=\frac{%
\omega _{\textrm{S}}}{Q_{\textrm{S}}}  \label{threshold kappa}
\end{equation}%
Since the time-averaged stored energy in the pump wave inside the cavity
depicted in Figure 2 is%
\begin{equation}
\left\langle U_{p}\right\rangle =\frac{1}{\mu _{0}}\left\vert \mathcal{B}%
_{p}\right\vert ^{2}\mathcal{A}_{\textrm{eff}}\mathcal{L}_{\textrm{eff}}
\end{equation}%
we conclude from (\ref{threshold kappa}) that the threshold pump power
needed for parametric oscillation is%
\begin{equation}
\left\langle U_{p}\right\rangle _{\textrm{threshold}}=\frac{1}{2}\frac{m\Omega
\omega _{\textrm{S}}\mathcal{L}_{\textrm{eff}}^{2}}{Q_{\textrm{S}}Q_{\Omega }}
\label{SRS threshold}
\end{equation}%
This result is to be compared with the threshold pump power needed for
parametric oscillation for exciting an elastic mode of a mirror of a
Fabry-Perot resonator obtained \ by Braginsky and co-workers \cite%
{braginsky01}%
\begin{equation}
\left\langle U_{p}\right\rangle _{\textrm{Braginsky}}=\frac{1}{2}\frac{m\omega
_{s}^{2}L^{2}}{Q_{i}Q_{s}}  \label{Braginsky}
\end{equation}%
where $m$ is the mass of the mirror, $\omega _{s}$ is the frequency of the
elastic mode, $L$ is the length of the Fabry-Perot resonator, $Q_{i}$ is the
quality factor of a down-shifted \textquotedblleft idler\textquotedblright\
optical mode of the resonator, and $Q_{s}$ is the quality factor of the
elastic mode. By inspection of (\ref{SRS threshold}) and (\ref{Braginsky}),
we see that these thresholds are quite similar.

However, the electrodynamic $Q$ factor of SC microwave cavities is typically
on the order of $10^{10}$ \cite{kuhr07}, whereas the typical mechanical $Q$
factor for the best opto-mechanical oscillators, which are composed of
non-SC materials in the ongoing opto-mechanical experiments, is at most on
the order of $10^{5}$ \cite{aspelmeyer}. Therefore the question naturally
arises whether it is possible to replace these low-$Q$, non-SC mechanical
oscillators, with high-$Q$ SC mechanical oscillators, in which their
mechanical $Q$ can approach the typical electrodynamic $Q\sim 10^{10}$ of SC
microwave cavities. \ 

\begin{figure}[h]
%\sidecaption[t]
\centering
\includegraphics[angle=0,width=.8%
\textwidth]{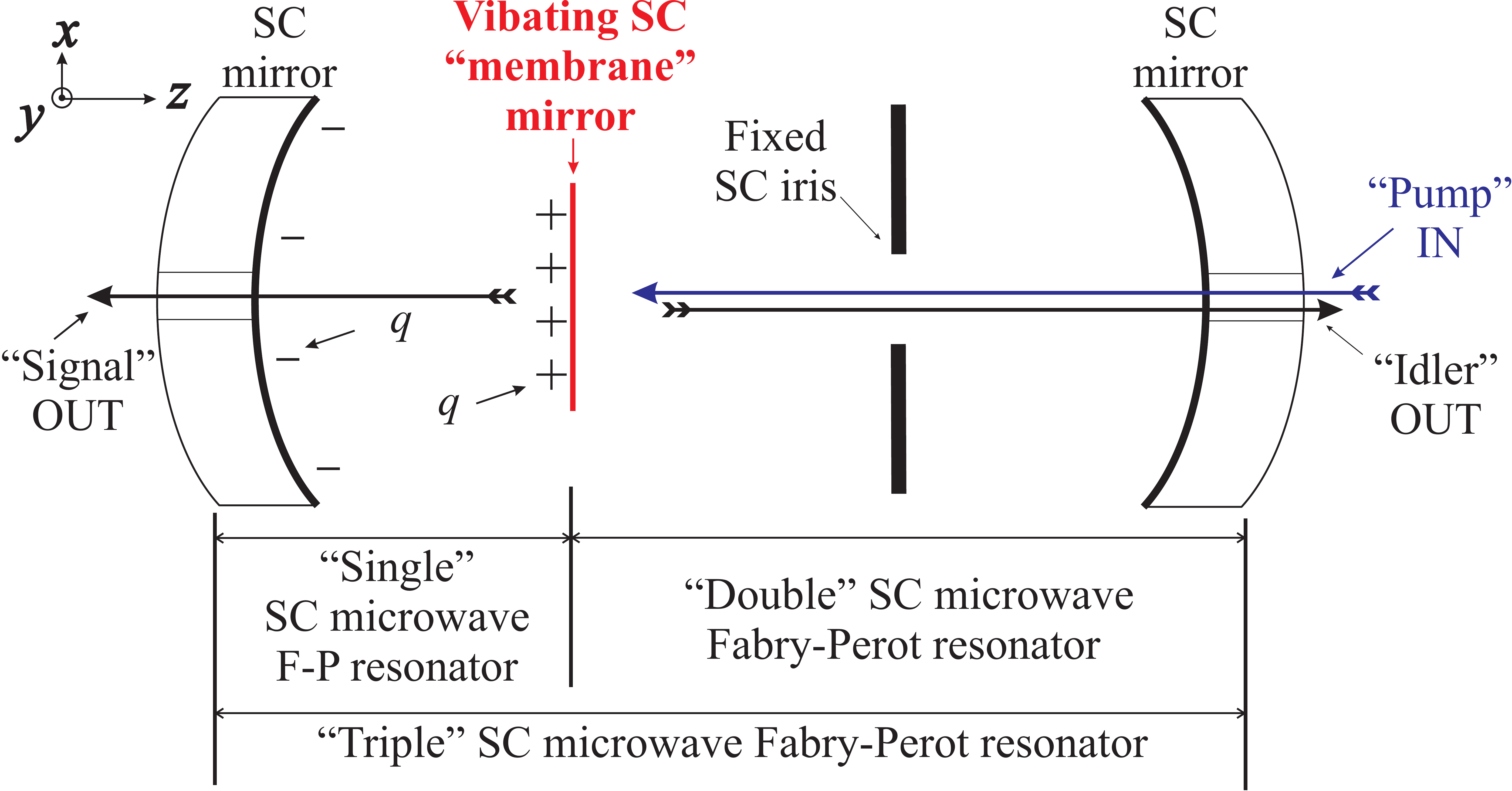} 
%\label{fig:REFERENCE}
\caption{(Color online) A \textquotedblleft triple\textquotedblright\ SC
microwave Fabry-Perot cavity \protect\cite{FQMT11}\ consists of a
\textquotedblleft single\textquotedblright\ SC cavity separated by a
vibrating SC \textquotedblleft membrane\textquotedblright\ (in red)\ from a
\textquotedblleft double\textquotedblright\ SC cavity with a fixed SC iris
(in black) at its center. This separating membrane is totally \emph{%
impermeable} to all microwaves. The membrane is electrostatically charged on
it left surface with a charge $+q$, and the left SC mirror is charged on its
right surface with a charge $-q$. \textquotedblleft Pump\textquotedblright\
microwaves (in blue) enter into the system through the right hole, and
\textquotedblleft signal\textquotedblright\ waves leave the system through
the left hole, but \textquotedblleft idler\textquotedblright\ waves leave
the system through the right hole.}
\end{figure}

%\FRAME{fhFU}{4.8551in}{2.5417in}{0pt}{\Qcb{A \textquotedblleft
%triple\textquotedblright\ SC microwave Fabry-Perot cavity \protect\cite%
%{FQMT11}\ consists of a \textquotedblleft single\textquotedblright\ SC
%cavity separated by a vibrating SC \textquotedblleft
%membrane\textquotedblright\ (in red)\ from a \textquotedblleft
%double\textquotedblright\ SC cavity with a fixed SC iris (in black) at its
%center. This separating membrane is totally \emph{impermeable} to all
%microwaves. The membrane is electrostatically charged on it left surface
%with a charge $+q$, and the left SC mirror is charged on its right surface
%with a charge $-q$. \textquotedblleft Pump\textquotedblright\ microwaves (in
%blue) enter into the system through the right hole, and \textquotedblleft
%signal\textquotedblright\ waves leave the system through the left hole, but
%\textquotedblleft idler\textquotedblright\ waves leave the system through
%the right hole.}}{}{triple-fp-with-vibrating-wire-mirror--12-28-12.eps}{%
%\special{language "Scientific Word";type "GRAPHIC";maintain-aspect-ratio
%TRUE;display "USEDEF";valid_file "F";width 4.8551in;height 2.5417in;depth
%0pt;original-width 22.409in;original-height 11.7in;cropleft "0";croptop
%"1";cropright "1";cropbottom "0";filename
%'../Triple-FP-with-vibrating-wire-mirror--12-28-12.eps';file-properties
%"XNPEU";}}

One possible answer to this question is the \textquotedblleft triple SC
microwave Fabry-Perot resonator\textquotedblright\ shown in Figure 4 \cite%
{FQMT11}, in which a \emph{charged} SC membrane is extremely tightly coupled
via its electrostatic charge to the longitudinal electric fields of a
transverse magnetic mode of a high $Q$ SC microwave cavity. The idea here is
that when the charge on the SC membrane is sufficiently large, then the
mechanical dynamics of the membrane will be \textquotedblleft
slaved\textquotedblright\ to follow closely the electromagnetic dynamics of
the high-$Q$ microwave SC mode. Calculations \cite{slaved dynamics of
membrane}\ show that one only needs a charge of pico-Coulombs for this to
happen. Note that as a result of the \textquotedblleft
slaved\textquotedblright\ dynamics, the SC membrane will be moving at
microwave, and not at acoustical, frequencies. This means that the motion of
the membrane will be essentially that of a \textquotedblleft
free\textquotedblright\ mass, which is being driven solely by Maxwell's
stress tensor. Therefore this microwave-frequency motion will be independent
of the elastic and dissipative mechanical properties of the membrane.

Another important feature of the configuration shown in Figure 4 is that the
signal and the idler waves are \emph{spatially\ separated} into two
disjoint, high $Q$ SC cavities, which are separated from each other by a
common, vibrating SC membrane. It turns out that this leads to two \emph{%
separate} $Q$ factors in its denominator of the threshold for parametric
oscillation, which arises due to this separation. We shall therefore call
the parametric oscillator configuration of Figure 4 a \textquotedblleft
separated parametric oscillator,\textquotedblright\ in contrast to that in
Figure 2, which we shall call an \textquotedblleft unseparated parametric
oscillator.\textquotedblright\ The threshold for the separated parametric
oscillator of Figure 4 will turn out to be at least a factor of $10^{5}$
lower than that of the unseparated parametric oscillator of Figure 2.

We start the analysis of the separated parametric amplifier depicted in
Figure 4 by examining the work done by the \textquotedblleft
pump\textquotedblright\ wave on the moving SC \textquotedblleft
membrane,\textquotedblright\ when it produces a displacement by an amount $%
\Delta z$ of the membrane to the left along the axis of the
\textquotedblleft double\textquotedblright\ Fabry-Perot resonator on the
right side of the membrane. The work done during this displacement is%
\begin{equation}
\Delta W=\left( \frac{1}{2\mu _{0}}B^{2}\right) \cdot \mathcal{A}_{\textrm{eff}%
}\Delta z  \label{work done by moving membrane}
\end{equation}%
where $\mathcal{A}_{\textrm{eff}}$ is the effective hemiconfocal spot size at
the membrane, $\Delta z$ is the displacement of this membrane, and 
\begin{equation}
u_{B}=\frac{1}{2\mu _{0}}B^{2}  \label{energy density of magnetic field}
\end{equation}%
is the energy density of the magnetic field evaluated at the right surface
of the \textquotedblleft membrane,\textquotedblright\ which is the pressure
arising from the Maxwell stress tensor (\ref{Tij for magnetic case}), i.e.,
a pressure being exerted upon the membrane that can cause a change of the
volume inside the \textquotedblleft double\textquotedblright\ Fabry-Perot
resonator on the right side of the membrane%
\begin{equation}
\Delta V=\mathcal{A}_{\textrm{eff}}\Delta z  \label{volume change on left side}
\end{equation}%
where $\mathcal{A}_{\textrm{eff}}$ is the effective area of the membrane,
which is determined by the hemiconfocal spot size of the mode on the right
side of the membrane. For simplicity, we shall assume here that the
hemiconfocal spot size of the mode on the left side of the membrane also has
the same $\mathcal{A}_{\textrm{eff}}$.

The instantaneous mechanical work $\Delta W$ in (\ref{work done by moving
membrane}) done by the \textquotedblleft pump\textquotedblright\ upon the
fields of the resonator can be rewritten in the form%
\begin{equation}
\Delta W=P\Delta V  \label{PdV work done by piston}
\end{equation}%
where the instantaneous pressure $P$ on the \textquotedblleft
membrane\textquotedblright\ is%
\begin{equation}
P=\frac{1}{2\mu _{0}}B^{2}  \label{pressure on plate}
\end{equation}%
which is equal to the instantaneous energy density $u_{B}$ in (\ref{energy
density of magnetic field}) evaluated at the right surface of the membrane.\
It is clear from the expression for the work in (\ref{PdV work done by
piston}) that $\Delta W$ can be interpreted as if it were\ the work being
done by a moving piston acting on a thermodynamic system, here, the
radiation fields inside a cavity.

Now we shall presently see that if energy were to be continually supplied to
the \textquotedblleft double Fabry-Perot\textquotedblright\ resonator on the
right side of the membrane\ by some continuous-wave, external microwave pump
waveform oscillating at a frequency $\omega _{p}$ (i.e., the
\textquotedblleft pump\textquotedblright\ frequency) entering through the
right hole of Figure 4, then the exponential amplification of some seed
\textquotedblleft signal\textquotedblright\ waveform at a frequency of $%
\omega _{s}$ within the \textquotedblleft single\textquotedblright\
Fabry-Perot on the left side of the membrane,\ simultaneously with the
exponential amplification of some seed \textquotedblleft
idler\textquotedblright\ waveform at a frequency of $\omega _{i}$ within the
\textquotedblleft double\textquotedblright\ Fabry-Perot on the right side of
the membrane,\ can occur. This amplification effect can arise from the
mutual reinforcement of the signal and idler waves at the expense of the pump
wave, in which the pump wave beats with the idler wave via the Maxwell
stress tensor to produce more of the signal wave, and the signal wave
modulates the pump wave via the Doppler effect to produce more of the idler
wave, etc. The mutual reinforcement of the two \textquotedblleft
seed\textquotedblright\ waves will lead to an instability above a certain
threshold, i.e., to the parametric oscillation of both the signal and idler
waves that produces macroscopic amounts of both kinds of waves, which then
leave the system in opposite directions via the left hole and the right hole
of Figure 4, respectively, just like in a laser.

For parametric amplification to occur, the frequency-matching condition%
\begin{equation}
\omega _{p}=\omega _{s}+\omega _{i}  \label{pump frequency = signal + idler}
\end{equation}%
must be satisfied. The meaning of the relationship can be most easily seen
by multiplying it by the Planck's constant $\hbar $ so that one obtains the
relationship%
\begin{equation}
\hbar \omega _{p}=\hbar \omega _{s}+\hbar \omega _{i}
\label{photon of pump = photon of signal + photon of idler}
\end{equation}%
In other words, in the parametric amplification process, one signal photon $%
\hbar \omega _{s}$ is simultaneously created along with one idler photon $%
\hbar \omega _{i}$ at the expense of one pump photon $\hbar \omega _{p}$,
which is annihilated during this \textquotedblleft photon pair-creation
process.\textquotedblright\ In this process, an entangled pair of signal and
idler photons, with the signal photon appearing on the left side, and the
idler photon appearing on the right side of the membrane, respectively, will
be produced in a correlated emission event inside the \textquotedblleft
triple\textquotedblright\ Fabry-Perot resonator depicted in Figure 4.\ This
photon pair-creation process is described by the interaction Hamiltonian%
\begin{equation}
H_{\textrm{int}}\propto a_{p}a_{s}^{\dag }a_{i}^{\dag }\textrm{ + hermitian
adjoint}
\end{equation}%
which is a generator of a \emph{two-mode} squeezed state \cite%
{garrison&chiao}.

Let the microwave pump magnetic field just outside of the right surface of
the membrane have the form%
\begin{equation}
B_{p}=\mathcal{B}_{p}\exp \left( -i\omega _{p}t\right) +\textrm{c.c.}
\label{pump magnetic field waveform}
\end{equation}%
where the pump magnetic field vector points transversely to the membrane
immediately outside of its right surface, with $\mathcal{B}_{p}$ being the
complex amplitude of the pump magnetic field.

Similarly, let the \textquotedblleft seed\textquotedblright\ idler magnetic
field just outside of the right surface of the membrane have the form%
\begin{equation}
B_{i}=\mathcal{B}_{i}\exp \left( -i\omega _{i}t\right) +\textrm{c.c.}
\label{idler magnetic field waveform}
\end{equation}%
which is a vector parallel to the magnetic field vector of the pump wave
immediately outside of its right surface, with $\mathcal{B}_{i}$ being the
complex amplitude of the idler magnetic field. We shall assume that the pump
is tuned to be on resonance with the upper member of the spectral doublet of
the \textquotedblleft double\textquotedblright\ Fabry-Perot (see Figure 3),
and that the idler is tuned to be on resonance with the lower member of this
doublet. However, the off-resonance, parasitic \textquotedblleft
anti-Stokes\textquotedblright\ (i.e., the Dopper up-shifted) frequency
component arising from the motion of the membrane will be suppressed, and
hence neglected.

To calculate the coefficient of parametric amplification, let us assume that
the pump wave is much stronger than both a very weak \textquotedblleft
seed\textquotedblright\ idler wave and a very weak \textquotedblleft
seed\textquotedblright\ signal wave, so that $\left\vert \mathcal{B}%
_{p}\right\vert >>\left\vert \mathcal{B}_{i}\right\vert $ and $\left\vert 
\mathcal{B}_{p}\right\vert >>\left\vert \mathcal{B}_{s}\right\vert $. It
follows from (\ref{pump frequency = signal + idler}) and (\ref{idler
magnetic field waveform}) that the square of the total magnetic field
evaluated at the right surface of the \textquotedblleft
membrane\textquotedblright\ will have the form

\begin{equation}
B^{2}=\left( B_{p}+B_{i}\right) ^{2}=(\mathcal{B}_{p}\exp \left( -i\omega
_{p}t\right) +\mathcal{B}_{i}\exp \left( -i\omega _{i}t\right) +\textrm{c.c.}%
)^{2}  \label{B squared}
\end{equation}%
If we define the \textquotedblleft beat frequency\textquotedblright\ as%
\begin{equation}
\Omega =\omega _{p}-\omega _{i}
\end{equation}%
then we see that there will arise cross terms in the square of the magnetic
field (\ref{B squared}) which will contain terms that vary at the beat
frequency $\Omega $, viz.,%
\begin{eqnarray}
\left( B^{2}\right) _{\Omega } &=&\mathcal{B}_{p}\exp \left( -i\omega
_{p}t\right) \mathcal{B}_{i}^{\ast }\exp \left( +i\omega _{i}t\right) +\textrm{%
c.c.}  \nonumber \\
&=&\mathcal{B}_{p}\mathcal{B}_{i}^{\ast }\exp \left( -i\Omega t\right) +%
\textrm{c.c.}
\end{eqnarray}%
Therefore there will exist a pressure being exerted on the membrane that
varies at the beat frequency $\Omega $ of the form 
\begin{equation}
\left( P\right) _{\Omega }=\frac{1}{2\mu _{0}}\left( B^{2}\right) _{\Omega }=%
\frac{1}{2\mu _{0}}\left( \mathcal{B}_{p}\mathcal{B}_{i}^{\ast }\exp \left(
-i\Omega t\right) +\textrm{c.c.}\right)  \label{pressure at beat frequency}
\end{equation}%
If we define the complex pressure amplitude $\mathcal{P}_{\Omega }$ as
follows:%
\begin{equation}
\mathcal{P}_{\Omega }=\frac{1}{2\mu _{0}}\mathcal{B}_{p}\mathcal{B}%
_{i}^{\ast }  \label{complex pressure amplitude}
\end{equation}%
then the pressure exerted on the membrane which varies at the beat frequency 
$\Omega $ will have the form 
\begin{equation}
\left( P\right) _{\Omega }=\mathcal{P}_{\Omega }\exp \left( -i\Omega
t\right) +\textrm{c.c.}
\end{equation}

But the beat frequency $\Omega $ will be assumed to be tuned into resonance
with the signal frequency, i.e.,%
\begin{equation}
\Omega =\omega _{p}-\omega _{i}=\omega _{s}
\label{beat frequency = pump - idler = signal}
\end{equation}%
so that the membrane can be driven at the resonance frequency $\omega _{s}$
of the \textquotedblleft single\textquotedblright\ Fabry-Perot resonator to
the left of the membrane. Thus power from the right side of the membrane can
be fed resonantly by the motion of the membrane into the signal
\textquotedblleft seed\textquotedblright\ waveform on the left side.

Due to the presence of the electrostatic charge $+q$ on the left surface of
the membrane, and the relationship%
\begin{equation}
F_{z}(t)=qE_{z}\left( t\right)  \label{F_z = q E_z}
\end{equation}%
where $F_{z}(t)$ is the Coulomb force on the membrane exerted on the charge $%
+q$ by the longitudinal electric field $E_{z}\left( t\right) $ of a
transverse magnetic mode of the \textquotedblleft single\textquotedblright\
Fabry-Perot resonator on the left side of the membrane \cite{FQMT11}, it
follows that the motion of the membrane in the longitudinal $z$ direction
will be \textquotedblleft slaved\textquotedblright\ through $E_{z}\left(
t\right) $\ to the dynamics of this transverse magnetic mode \cite[Appendix
B, where it was shown that one only needs $q\approx 20$ pC for the Coulomb
force to dominate the dynamics of the membrane]{FQMT11}. This is due to the
tight coupling between $\Delta z(t)$ and $E_{z}\left( t\right) $\ which
arises from the electrostatic charge $+q$ \cite{slaved dynamics of membrane}.

Hence let us introduce an Ansatz that the displacement of the membrane has
the form%
\begin{equation}
\left( \Delta z\left( t\right) \right) _{\Omega }=\varepsilon _{\Omega
}\left( t\right) \exp \left( -i\Omega t\right) +\textrm{c.c.}
\label{displacement of membrane}
\end{equation}%
where $\varepsilon _{\Omega }\left( t\right) $ is some \emph{slowly-varying}
complex displacement amplitude of the membrane, which is the slowly-varying
envelope of the \emph{fast} beat frequency phase factor $\exp \left(
-i\Omega t\right) $.

Hence the velocity of the membrane will have the form 
\begin{equation}
\left( v\right) _{\Omega }=\frac{d\left( \Delta z\right) _{\Omega }}{dt}%
\approx \nu _{\Omega }\left( t\right) \exp \left( -i\Omega t\right) +\textrm{%
c.c.}  \label{velocity at beat frequency}
\end{equation}%
where the \emph{slowly-varying} complex velocity amplitude of the moving
membrane is%
\begin{equation}
\nu _{\Omega }\left( t\right) \approx -i\Omega \varepsilon _{\Omega }\left(
t\right)  \label{velocity of membrane}
\end{equation}%
within the \textquotedblleft slowly varying envelope
approximation\textquotedblright\ \cite{SVEA}, and the acceleration of the
membrane will have the form%
\begin{equation}
\left( a\right) _{\Omega }=\left( \frac{dv}{dt}\right) _{\Omega }\approx
\alpha _{\Omega }\left( t\right) \exp \left( -i\Omega t\right) +\textrm{c.c.}
\end{equation}%
where the \emph{slowly-varying} complex acceleration amplitude of the moving
membrane is%
\begin{equation}
\alpha _{\Omega }\left( t\right) \approx -\Omega ^{2}\varepsilon _{\Omega
}\left( t\right)  \label{acceleration of membrane}
\end{equation}%
also within the slowly varying envelope approximation.

The force due to the pressure (\ref{pressure at beat frequency}) being
exerted on the membrane will have the form%
\begin{equation}
\left( F\right) _{\Omega }=\mathcal{F}_{\Omega }\left( t\right) \exp \left(
-i\Omega t\right) +\textrm{c.c.}
\end{equation}%
where the \emph{slowly-varying} complex force amplitude acting on the
membrane is%
\begin{equation}
\mathcal{F}_{\Omega }\left( t\right) =\frac{1}{\mu _{0}}\mathcal{B}_{p}%
\mathcal{B}_{i}^{\ast }\left( t\right) \mathcal{A}_{\textrm{eff}}
\label{complex force amplitude}
\end{equation}%
where $\mathcal{A}_{\textrm{eff}}$ is the effective area of the membrane, and
where, in the \textquotedblleft undepleted pump\textquotedblright\
approximation \cite{Boyd}, we have assumed that the pump amplitude $\mathcal{%
B}_{p}$ is independent of time, but that the idler amplitude $\mathcal{B}%
_{i}^{\ast }\left( t\right) $ can be a slowly varying function of time due
to its amplification.

Then the time-averaged mechanical power fed into the membrane's motion from
the radiation, and hence into the signal wave of the \textquotedblleft
single\textquotedblright\ Fabry-Perot cavity of Figure 4, using (\ref%
{velocity of membrane}) and (\ref{complex force amplitude}), is%
\begin{eqnarray}
\left\langle \frac{dW}{dt}\right\rangle _{\textrm{signal}} &=&\left\langle
F\cdot v\right\rangle  \nonumber \\
&=&\left\langle \mathcal{F}_{\Omega }\cdot \nu _{\Omega }^{\ast }+\textrm{c.c.}%
\right\rangle =\left\langle \mathcal{F}_{\Omega }\cdot i\Omega \varepsilon
_{\Omega }^{\* }+\textrm{c.c.}\right\rangle  \nonumber \\
&=&-2\textrm{Im}\left( \mathcal{F}_{\Omega }\Omega \varepsilon _{\Omega }^{\ast
}\right) =-2\textrm{Im}\left( \frac{1}{\mu _{0}}\mathcal{B}_{p}\mathcal{B}_{i}^{\ast
}\left( t\right) \mathcal{A}_{\textrm{eff}}\cdot \Omega \varepsilon _{\Omega
}^{\ast }\left( t\right) \right)  \label{time-averaged power fed to membrane}
\end{eqnarray}%
Let the complex amplitudes of the pump, idler, and signal waveforms have the
complex polar forms%
\begin{eqnarray}
\mathcal{B}_{p} &=&\left\vert \mathcal{B}_{p}\right\vert \exp \left( i\phi
_{p}\right)  \label{pump polar form} \\
\mathcal{B}_{i} &=&\left\vert \mathcal{B}_{i}\right\vert \exp \left( i\phi
_{i}\right)  \label{idler polar form} \\
\varepsilon _{\Omega } &=&\left\vert \varepsilon _{\Omega }\right\vert \exp
\left( i\phi _{s}\right)  \label{signal polar form}
\end{eqnarray}%
By inspection of (\ref{time-averaged power fed to membrane}), we see that
the maximum power transfer from the radiation fields into the membrane's
motion occurs when the phases of pump, signal, and idler waveforms (i.e., (%
\ref{pump polar form}), (\ref{idler polar form}), and (\ref{signal polar
form})) are adjusted so as to satisfy the condition%
\begin{equation}
\phi _{p}-\phi _{i}-\phi _{s}=-\frac{\pi }{2}  \label{phase condition}
\end{equation}%
whereupon the maximum mechanical power fed into the membrane becomes%
\begin{equation}
\left\langle \frac{dW}{dt}\right\rangle _{\max \textrm{, signal}}=+\frac{2%
\mathcal{A}_{\textrm{eff}}\Omega }{\mu _{0}}\left\vert \mathcal{B}%
_{p}\right\vert \left\vert \mathcal{B}_{i}\left( t\right) \right\vert
\left\vert \varepsilon _{\Omega }\left( t\right) \right\vert
\label{maximum power transfer to membrane}
\end{equation}

Neglecting for the moment all dissipative losses, the kinetic energy of the
membrane must \emph{grow} due to this \emph{positive} mechanical power being
fed into it. Hence, invoking the principle of the conservation of energy, we
get the equation%
\begin{equation}
\left\langle \frac{dW}{dt}\right\rangle _{\max \textrm{, signal}}=\frac{d}{dt}%
\left( \frac{1}{2}m\left\langle v^{2}\right\rangle \right) =\frac{d}{dt}%
\left( m\Omega ^{2}\left\vert \varepsilon _{\Omega }\left( t\right)
\right\vert ^{2}\right)
\end{equation}%
where we have used (\ref{velocity of membrane}) and the fact that $%
\left\langle v^{2}\right\rangle =2\left\vert \nu _{\Omega }\right\vert ^{2}$%
. Therefore%
\begin{equation}
\frac{d}{dt}\left( m\Omega ^{2}\left\vert \varepsilon _{\Omega }\right\vert
^{2}\right) =2m\Omega ^{2}\left\vert \varepsilon _{\Omega }\right\vert \frac{%
d\left\vert \varepsilon _{\Omega }\right\vert }{dt}=\frac{2\mathcal{A}_{%
\textrm{eff}}\Omega }{\mu _{0}}\left\vert \mathcal{B}_{p}\right\vert
\left\vert \mathcal{B}_{i}\right\vert \left\vert \varepsilon _{\Omega
}\right\vert
\end{equation}%
We thus arrive at an ODE for the rate of growth of the magnitude $\left\vert
\varepsilon _{\Omega }\right\vert $\ of the displacement of the membrane%
\begin{equation}
\frac{d\left\vert \varepsilon _{\Omega }\right\vert }{dt}=\frac{\mathcal{A}_{%
\textrm{eff}}}{\mu _{0}m\Omega }\left\vert \mathcal{B}_{p}\right\vert
\left\vert \mathcal{B}_{i}\right\vert  \label{ODE for displacement magnitude}
\end{equation}

Next, we shall obtain a similar ODE for the rate of growth of the magnitude $%
\left\vert \mathcal{B}_{i}\right\vert $\ of the idler wave. We start from
the motional EMF created by the motion of the vibrating SC membrane, which
leads to the generation of the motional electric field

\begin{equation}
\mathbf{E}=\mathbf{v\times B}  \label{E'=-vxB}
\end{equation}%
This relationship implies that the sinusoidal, back-and-forth motion of the
mirror at a frequency $\Omega $ will be modulating the magnetic field
oscillating at the pump frequency $\omega _{p}$, such that an idler electric
field at the surface of the mirror oscillating at the idler frequency $%
\omega _{i}$ will be generated, i.e.,%
\begin{equation}
\left( \mathbf{E}\right) _{i}=\left( \mathbf{v}\right) _{\Omega }\mathbf{%
\times }\left( \mathbf{B}\right) _{p}  \label{E' at idler produced}
\end{equation}%
This is a manifestation of the Doppler effect, in which the sinusoidal
motion of the mirror will produce Stokes and anti-Stokes sidebands around
the pump frequency $\omega _{p}$. However, due to the doublet spectrum
depicted in Figure 3 in the \textquotedblleft double\textquotedblright\
Fabry-Perot resonator, only the down-shifted, first-order Stokes sideband
will be resonant with the resonator. Therefore we shall neglect henceforth
the anti-Stokes sideband, and all the other higher order Doppler sidebands.

Note that when the membrane is moving towards the left in Figure 4, which is
the direction in which the magnetic pressure due to the pump wave is
pushing, this pressure will deliver power into the motion of the moving
membrane, and \emph{simultaneously}, will deliver power into the
red-shifted, first-order Doppler sideband, i.e., the idler wave, via the
relationship (\ref{E' at idler produced}). This will lead to parametric
amplification of the membrane's motion.

Using the Cartesian coordinate system shown in Figure 4, we shall assume
that the instantaneous velocity of the mirror is pointing in the $-z$
direction, and that the instantaneous pump magnetic field vector $\mathbf{B}%
_{p}$ is pointing in $+x\,\ $direction, so that the instantaneous motional $%
\mathbf{E}$ field will be pointing in the $+y$ direction. Thus, in terms of
the complex amplitudes, (\ref{E' at idler produced}) reduces down to%
\begin{equation}
\mathcal{E}_{i}=\nu _{\Omega }^{\ast }\mathcal{B}_{p}
\label{complex idler = velocity at signal x complex pump}
\end{equation}

Now the time-averaged power delivered into the idler wave by the motional $%
\mathbf{E}$ field acting on the current density $\mathbf{j}$ induced by the
idler wave at the right surface of the SC mirror, is%
\begin{equation}
\left\langle \frac{dW}{dt}\right\rangle _{\textrm{idler}}=\int \left\langle 
\mathbf{j\cdot E}\right\rangle dV=\left( j_{i}^{\ast }\mathcal{E}_{i}\right) 
\mathcal{A}_{\textrm{eff}}\delta +\textrm{c.c.}  \label{j.E' power}
\end{equation}%
where $j_{i}^{\ast }$ is the complex conjugate of the supercurrent density
amplitude flowing on the surface of the mirror at the idler frequency $%
\omega _{i}$, $\mathcal{E}_{i}$ is the complex motional electric field
amplitude at $\omega _{i}$, $\mathcal{A}_{\textrm{eff}}$ is the effective\
focal area of the membrane, and $\delta $ is the London penetration depth,
within which the supercurrents $j_{i}^{\ast }$ will be flowing near the
surface of the membrane. To calculate $j_{i}^{\ast }$ in (\ref{j.E' power}),
we use Ampere's circuital law and a rectangular loop straddling the surface
of the SC mirror, to get%
\begin{equation}
\mathcal{B}_{i}^{\ast }=\mu _{0}j_{i}^{\ast }\delta
\label{idler magnetic amplitude}
\end{equation}%
where $\delta $ is London's penetration depth. Solving for $j_{i}^{\ast }$
from (\ref{idler magnetic amplitude}), and substituting it into (\ref{j.E'
power}) using (\ref{complex idler = velocity at signal x complex pump}), we
get%
\begin{eqnarray}
\left\langle \frac{dW}{dt}\right\rangle _{\textrm{idler}} &=&\frac{1}{\mu _{0}}%
\left( \nu _{\Omega }^{\ast }\mathcal{B}_{p}\mathcal{B}_{i}^{\ast }\right) 
\mathcal{A}_{\textrm{eff}}+\textrm{c.c.}  \nonumber \\
&=&\frac{1}{\mu _{0}}\left( \left( i\Omega \varepsilon _{\Omega }^{\ast
}\right) \mathcal{B}_{p}\mathcal{B}_{i}^{\ast }\right) \mathcal{A}_{\textrm{eff%
}}+\textrm{c.c.}
\end{eqnarray}%
where in the last step we used (\ref{velocity of membrane}) for the complex
conjugate of the complex velocity amplitude $\nu _{\Omega }^{\ast }$ of the
membrane.

To maximize the power transferred to the idler, we again choose the phase
condition (\ref{phase condition}) between the pump, idler and signal complex
amplitudes, and find%
\begin{equation}
\left\langle \frac{dW}{dt}\right\rangle _{\textrm{max, idler}}=+\frac{2}{\mu
_{0}}\Omega \left\vert \varepsilon _{\Omega }\right\vert \left\vert \mathcal{%
B}_{p}\right\vert \left\vert \mathcal{B}_{i}\right\vert \mathcal{A}_{\textrm{%
eff}}  \label{max power to idler}
\end{equation}

Assuming the absence of all dissipation, and invoking once again the
principle of the conservation of energy, but this time for the idler wave,
we obtain%
\begin{eqnarray}
\left\langle \frac{dW}{dt}\right\rangle _{\textrm{max, idler}} &=&\frac{d}{dt}%
\left( \frac{1}{2\mu _{0}}\left\langle B_{i}^{2}\right\rangle \right) 
\mathcal{A}_{\textrm{eff}}\mathcal{L}_{\textrm{eff}}=\frac{d}{dt}\left( \frac{1}{%
\mu _{0}}\left\vert \mathcal{B}_{i}\right\vert ^{2}\right) \mathcal{A}_{%
\textrm{eff}}\mathcal{L}_{\textrm{eff}}  \nonumber \\
&=&+\frac{2}{\mu _{0}}\Omega \left\vert \varepsilon _{\Omega }\right\vert
\left\vert \mathcal{B}_{p}\right\vert \left\vert \mathcal{B}_{i}\right\vert 
\mathcal{A}_{\textrm{eff}}  \label{conservation of energy for idler wave}
\end{eqnarray}%
where $\mathcal{L}_{\textrm{eff}}$ is the effective length of the
\textquotedblleft double\textquotedblright\ Fabry-Perot resonator. We thus
arrive at an ODE for the rate of growth of the idler wave%
\begin{equation}
\frac{d}{dt}\left\vert \mathcal{B}_{i}\right\vert =\frac{\Omega }{\mathcal{L}%
_{\textrm{eff}}}\left\vert \varepsilon _{\Omega }\right\vert \left\vert 
\mathcal{B}_{p}\right\vert =K_{2}\left\vert \varepsilon _{\Omega }\right\vert
\label{ODE for idler}
\end{equation}%
where the constant of proportionality $K_{2}$ is%
\begin{equation}
K_{2}=\frac{\Omega }{\mathcal{L}_{\textrm{eff}}}\left\vert \mathcal{B}%
_{p}\right\vert  \label{K2}
\end{equation}%
Note that this is of the same form as the ODE for the rate of growth of the
signal wave \cite{slaved dynamics of membrane} obtained earlier in (\ref{ODE
for displacement magnitude}), viz.,%
\begin{equation}
\frac{d}{dt}\left\vert \varepsilon _{\Omega }\right\vert =\frac{\mathcal{A}_{%
\textrm{eff}}}{\mu _{0}m\Omega }\left\vert \mathcal{B}_{p}\right\vert
\left\vert \mathcal{B}_{i}\right\vert =K_{1}\left\vert \mathcal{B}%
_{i}\right\vert  \label{ODE for signal}
\end{equation}%
where the constant of proportionality $K_{1}$ is%
\begin{equation}
K_{1}=\frac{\mathcal{A}_{\textrm{eff}}}{\mu _{0}m\Omega }\left\vert \mathcal{B}%
_{p}\right\vert  \label{K1}
\end{equation}%
This implies that there exists a mutual enhancement of the signal and idler
waves that leads to their \emph{exponential} growth, i.e., to the \emph{%
parametric amplification} of both waves. To see this, let us rewrite the two
equations (\ref{ODE for idler}) and (\ref{ODE for signal}) in the following $%
2\times 2$ matrix form:%
\begin{equation}
\frac{d}{dt}\left( 
\begin{array}{c}
\left\vert \varepsilon _{\Omega }\right\vert \\ 
\left\vert \mathcal{B}_{i}\right\vert%
\end{array}%
\right) =\left( 
\begin{array}{cc}
0 & K_{1} \\ 
K_{2} & 0%
\end{array}%
\right) \left( 
\begin{array}{c}
\left\vert \varepsilon _{\Omega }\right\vert \\ 
\left\vert \mathcal{B}_{i}\right\vert%
\end{array}%
\right) =\Lambda \left( 
\begin{array}{c}
\left\vert \varepsilon _{\Omega }\right\vert \\ 
\left\vert \mathcal{B}_{i}\right\vert%
\end{array}%
\right)  \label{2x2 matrix ODE}
\end{equation}%
where $\Lambda $ is the eigenvalue of the $2\times 2$ matrix, viz.,%
\begin{equation}
\Lambda =\pm \sqrt{K_{1}K_{2}}=\pm \sqrt{\frac{\mathcal{A}_{\textrm{eff}%
}\left\vert \mathcal{B}_{p}\right\vert ^{2}}{\mu _{0}m\mathcal{L}_{\textrm{eff}%
}}}  \label{solution for eigenvalue of growth}
\end{equation}%
The solution of (\ref{2x2 matrix ODE}) is%
\begin{equation}
\left( 
\begin{array}{c}
\left\vert \varepsilon _{\Omega }\right\vert \\ 
\left\vert \mathcal{B}_{i}\right\vert%
\end{array}%
\right) =\left( 
\begin{array}{c}
\left\vert \varepsilon _{\Omega }\right\vert \\ 
\left\vert \mathcal{B}_{i}\right\vert%
\end{array}%
\right) _{t=0}\exp \left( \Lambda t\right)
\end{equation}%
The meaning of the \emph{positive} root for $\Lambda $ is that it represents
the rate of exponential \emph{growth} of the amplitudes of the coupled
signal and idler waves, when the phase condition (\ref{phase condition}), $%
\phi _{p}-\phi _{i}-\phi _{s}=-\pi /2$, for maximum power \emph{delivery} to
these coupled waves, is satisfied, whereas the \emph{negative} root for $%
\Lambda $ is that it represents the rate of exponential \emph{decay} of the
amplitudes of the coupled signal and idler waves, when the anti-phase
condition, $\phi _{p}-\phi _{i}-\phi _{s}=+\pi /2\,$,\ for maximum power 
\emph{extraction} from these coupled waves, is satisfied. Whether one gets
exponential growth or exponential decay of the waves thus depends on the
choices of the initial phases of the pump, signal, and idler waves. This
kind of \emph{phase-dependent} amplification is the signature of the
production of a \emph{squeezed state} of the vacuum.

Next, let us introduce a dissipative loss phenomenologically into the ODE
for the idler (\ref{ODE for idler}) as follows:%
\begin{equation}
\frac{d}{dt}\left\vert \mathcal{B}_{i}\right\vert -\frac{2}{\tau _{i}}%
\left\vert \mathcal{B}_{i}\right\vert =K_{2}\left\vert \varepsilon _{\Omega
}\right\vert  \label{ODE for idler with loss}
\end{equation}%
where $\tau _{i}$ is the \textquotedblleft cavity ring-down
time\textquotedblright\ for the energy stored in the idler cavity mode on
the right side of the membrane after the pump wave has been suddenly shut
off, and, similary, into the ODE for the signal (\ref{ODE for signal}) as
follows: 
\begin{equation}
\frac{d}{dt}\left\vert \varepsilon _{\Omega }\right\vert -\frac{2}{\tau _{s}}%
\left\vert \varepsilon _{\Omega }\right\vert =K_{1}\left\vert \mathcal{B}%
_{i}\right\vert  \label{ODE for signal with loss}
\end{equation}%
where $\tau _{s}$ is the \textquotedblleft cavity ring-down\
time\textquotedblright\ for the energy stored in the signal cavity mode on
the left side of the membrane after the pump wave has been suddenly shut
off. The relationships between the cavity ring-down times $\tau _{i}$ and $%
\tau _{s}$ of the two cavity modes, and their loaded quality factors $Q_{i}$
and $Q_{s}$, are%
\begin{eqnarray}
Q_{i} &=&\omega _{i}\tau _{i}\textrm{ }  \label{Q_i} \\
Q_{s} &=&\omega _{s}\tau _{s}  \label{Q_s}
\end{eqnarray}

At the threshold of parametric oscillation, there is a balance between gain
and loss such that there arises a steady-state situation in which%
\begin{equation}
\frac{d}{dt}\left\vert \mathcal{B}_{i}\right\vert =\frac{d}{dt}\left\vert
\varepsilon _{\Omega }\right\vert =0
\end{equation}%
Therefore, at threshold, the two ODE's (\ref{ODE for idler with loss}) and (%
\ref{ODE for signal with loss}) reduce down to the two algebraic equations%
\begin{eqnarray}
-\frac{2}{\tau _{i}}\left\vert \mathcal{B}_{i}\right\vert &=&K_{2}\left\vert
\varepsilon _{\Omega }\right\vert \\
-\frac{2}{\tau _{s}}\left\vert \varepsilon _{\Omega }\right\vert
&=&K_{1}\left\vert \mathcal{B}_{i}\right\vert
\end{eqnarray}%
Multiplying the left sides and the right sides of these two equations
together, we get%
\begin{equation}
\frac{4}{\tau _{i}\tau _{s}}=K_{2}K_{1}  \label{product of tau's and K's}
\end{equation}%
By using the relationships (\ref{K2}), (\ref{K1}), (\ref{Q_i}), and (\ref%
{Q_s}), we get from (\ref{product of tau's and K's}) the threshold condition%
\begin{equation}
\frac{4\omega _{i}\omega _{s}}{Q_{i}Q_{s}}=\frac{\mathcal{A}_{\textrm{eff}%
}\left\vert \mathcal{B}_{p}\right\vert ^{2}}{\mu _{0}m\mathcal{L}_{\textrm{eff}%
}}=\frac{\left\vert \mathcal{B}_{p}\right\vert ^{2}V_{\textrm{eff}}}{\mu _{0}m%
\mathcal{L}_{\textrm{eff}}^{2}}  \label{intermediate threshold condition}
\end{equation}%
Since the time-averaged stored energy stored in the pump cavity mode is%
\begin{equation}
\left\langle U_{p}\right\rangle =\frac{1}{2\mu _{0}}\left\langle
B_{p}^{2}\right\rangle V_{\textrm{eff}}=\frac{1}{\mu _{0}}\left\vert \mathcal{B%
}_{p}\right\vert ^{2}V_{\textrm{eff}}
\end{equation}%
we arrive from (\ref{intermediate threshold condition}) at the conclusion
that the threshold condition is%
\begin{equation}
\left\langle U_{p}\right\rangle _{\textrm{threshold}}=\frac{4m\omega
_{i}\omega _{s}\mathcal{L}_{\textrm{eff}}^{2}}{Q_{i}Q_{s}}
\label{U_p threshold =...}
\end{equation}%
This is to be compared with Braginski's threshold condition (\ref{Braginsky}%
) \cite{braginsky01}%
\begin{equation}
\left\langle U_{p}\right\rangle _{\textrm{threshold}}^{\textrm{Braginski}}=\frac{%
1}{2}\frac{m\omega _{s}^{2}L^{2}}{Q_{i}Q_{s}}
\end{equation}%
The above two expressions agree as to an order-of-magnitude estimate for the
threshold of parametric oscillation for the \textquotedblleft
triple\textquotedblright\ Fabry-Perot resonator configuration of Figure 4.

The required threshold input pump power $\left\langle \mathtt{P}%
_{p}\right\rangle _{\textrm{threshold}}$ for parametric oscillation due to
pump microwaves entering in through the right hole of the \textquotedblleft
triple\textquotedblright\ cavity configuration of Figure 4, can be found via
the steady-state condition%
\begin{equation}
\left\langle \mathtt{P}_{p}\right\rangle _{\textrm{threshold}}=\frac{1}{\tau
_{p}}\left\langle U_{p}\right\rangle _{\textrm{threshold}}
\end{equation}%
where the quality factor for the pump cavity mode $Q_{p}$ is related to the
pump cavity ring-down time $\tau _{p}$ by%
\begin{equation}
Q_{p}=\omega _{p}\tau _{p}
\end{equation}%
Finally, putting this together with (\ref{U_p threshold =...}), we conclude
that for parametric oscillation to occur in the configuration of Figure 4,
we need to inject a microwave pump power into the \textquotedblleft
triple\textquotedblright\ Fabry-Perot cavity the minimum amount of%
\begin{equation}
\left\langle \mathtt{P}_{p}\right\rangle _{\textrm{threshold}}=\frac{4m\omega
_{p}\omega _{i}\omega _{s}\mathcal{L}_{\textrm{eff}}^{2}}{Q_{p}Q_{i}Q_{s}}
\end{equation}

Numerically, if we assume that \cite{grid of wires} 
\begin{equation}
m=\textrm{ 2 mg}
\end{equation}%
\begin{equation}
\omega _{p}=2\pi \times \textrm{ 20 GHz}
\end{equation}%
\begin{equation}
\omega _{i}\approx \omega _{s}\approx 2\pi \times \textrm{10 GHz}
\end{equation}%
\begin{equation}
\mathcal{L}_{\textrm{eff}}\approx \lambda _{i}/2\approx \lambda _{s}/2\approx 
\textrm{3 cm}
\end{equation}%
\begin{equation}
Q_{p}\approx Q_{i}\approx Q_{s}\approx 10^{10}
\end{equation}%
then we conclude that we would require a microwave pump power at a frequency
of 20 GHz to be injected through the right hole of the \textquotedblleft
triple\textquotedblright\ Fabry-Perot cavity of at least%
\begin{equation}
\left\langle \mathtt{P}_{p}\right\rangle _{\textrm{threshold}}\approx \textrm{4 microwatts}
\end{equation}%
Above this minimum power level, there would result a parametric oscillation
effect in which a macroscopic amount of signal and idler microwaves centered
around 10 GHz, with powers on the order of microwatts (i.e., with
powers comparable to the pump threshold power), would be emitted in opposite
directions through the left and the right holes, respectively, of the
\textquotedblleft triple\textquotedblright\ Fabry-Perot cavity, like in a
laser. If the $Q$ is lowered by opening the outcoupling holes, the threshold will go up, but so will the output power of the parametric oscillator. For example, by lowering all the $Q$'s to $10^9$ instead of $10^{10}$, the threshold will be increased to 400 microwatts, but the output power of the dynamical Casimir effect will also increase by a few milliwatts. It should be stressed that the leftmost cavity, that is, the
\textquotedblleft single\textquotedblright\ Fabry-Perot resonator of Figure
4, is initially devoid of any radiation (i.e., it is initially an \emph{empty%
} cavity), so that the emission of a macroscopic amount of signal microwaves
through the left hole from the left side of the apparatus, would be a
dramatic manifestation of the dynamical Casimir effect, in which the
observed signal output must have built up exponentially starting\ solely
from vacuum fluctuations inside this initially empty resonator. Since the
dynamical Casimir effect is closely related to Hawking radiation according
to \cite{nation}, an observation of parametric oscillation resulting from
the moving SC membrane in Figure 4 would be a very interesting result from
the point of view of quantum field theory.

\section{The gravitational dynamical Casimir effect, and the generation of
coherent gravitational radiation}

In this final section, we speculate that the above ideas can be extended to
include the case of gravitational radiation. The physical concept that ties
all these ideas together is the crucial use of the DeWitt minimal coupling
rule in all of them.

In particular, we briefly comment on the possibility of extending the
\textquotedblleft separated parametric oscillator\textquotedblright\ idea
for generating EM microwaves by means of the vibrating SC membrane\ placed
inside the extremely high $Q$\textit{\ }\textquotedblleft
triple\textquotedblright\ SC cavity, as depicted in Figure 4, to the much
more speculative idea of generating GR microwaves using the same vibrating
SC membrane inside the same\textit{\ }\textquotedblleft
triple\textquotedblright\ SC cavity. This extension is based on the fact
that the interaction Hamiltonian $H_{\mathbf{h\cdot h}}$ in (\ref{H_h.h}) is
mathematically identical to that of the interaction Hamiltonian $H_{\mathbf{%
A\cdot A}}$\ in (\ref{H_A.A}). Furthermore, we are assuming that it is
permissible for gravitational radiation fields to be second quantized (see (%
\ref{[b,b dagger] = 1})).

However, for this extension of the parametric oscillator\ idea to work, it
is crucial that the walls SC cavity, including the surfaces of the moving SC
membrane, reflect GR microwaves with as high a reflectivity as in the case
of EM microwaves. In the paper \textquotedblleft Do mirrors for
gravitational waves exist?\textquotedblright\ \cite{physica e}, it was
predicted that even \emph{thin} SC films are highly reflective mirrors for
GR plane waves. This surprising prediction was based on the DeWitt minimal
coupling rule (\ref{DeWitt minimal coupling}) applied to the Ginzburg-Landau
theory of superconductivity. The \textquotedblleft off-diagonal long-range
order\textquotedblright\ (ODLRO) \cite{yang}\ nature of the Cooper pairs
causes these pairs to behave \emph{differently} from the ions in the ionic
lattice, for which ODLRO does not exist. As a result, inside the SC thin
film, the Cooper pairs, which exhibit constructive AB interference, \emph{do not} undergo geodesic motion, in contrast to
the ions, which \emph{do} undergo geodesic motion, in response to incident
GR radiation. This \emph{difference} in the internal motions of the Cooper
pairs and of the ions inside the SC in the presence of GR radiation, leads
to a charge separation effect induced by an incoming GR plane wave, such
that a huge back-action of the SC film on the GR wave that causes its
reflection, results.

If such SC mirrors for GR waves were indeed to exist in Nature, then moving
SC mirrors would not only be able to do work like a piston on these waves,
but would also simultaneously lead to a Doppler effect that leads to the
exponential amplification of these waves above the threshold for parametric
oscillation, as explained above. Thus, a laser-like generation of coherent
GR waves starting from vacuum fluctuations should become possible. If so, a
Hertz-like experiment for GR radiation at microwave frequencies \cite%
{Chiao-Townes-volume} would become feasible to perform.

\section{Acknowledgements}

DAS acknowledges the support of a 2012-2013 Fulbright Senior Scholar Grant.


\begin{thebibliography}{99}
\bibitem{newtonian limit} The constant $\varepsilon _{g}$ is determined by
the Newtonian limit of (\ref{Maxwell-like-eq-1}), which must yield Newton's
law of gravitation, with Newton's constant $G$ determined by Cavendish's
experiment.
\bibitem{mu_G} The constant $\mu _{g}$ is fixed by the fact that when
it is combined with the constant $\varepsilon _{g}$ to form the expression $%
1/\sqrt{\varepsilon _{g}\mu _{g}}$, one obtains the vacuum speed of
light $c$.
\bibitem{metric conventions} We follow the notation and sign conventions of
MTW \cite{MTW}, i.e., Greek indices denote spacetime indices running from 0
to 3; Latin indices denote spatial indices running from 1 to 3; repeated
indices are summed;$\ $the signature of the Minkowski metric is diag$%
(-1,+1,+1,+1)$.
\bibitem{DW-limits} It should be noted that the use of the time-dependent
gauge in which $\left( \mathbf{h}\left( \mathbf{r},t\right) \right)
_{i}=ch_{0i}\left( \mathbf{r},t\right) \neq 0$ in this context is only
considered for fields within matter, and not in vacuum, where the
time-dependent transverse-traceless gauge would be more appropriate.
\bibitem{grid of wires} Instead of a solid SC membrane, one could use a grid
of fine SC wires, for example, four fine SC wires arranged in a pattern
similar to the number sign \textquotedblleft \#\textquotedblright , in order
to reduce the mass, and thus the threshold.
\end{thebibliography}
\end{document}